\documentclass[12pt]{article}
\usepackage{bm,latexsym,amsmath,amssymb,amsfonts,mathrsfs}
\usepackage[dvipdfmx]{graphicx}
\usepackage[nosort]{cite}
\usepackage{color}
\usepackage{multirow}
\usepackage{enumitem}

\input{colordvi.tex}

\usepackage[hdivide={20mm,,20mm}, vdivide={25mm,,25mm}, nohead]{geometry}

\makeatletter
\@addtoreset{equation}{section}
\makeatother

\newcommand{\Ga}{{\Gamma}}
\newcommand{\De}{{\Delta}}
\newcommand{\Lm}{{\Lambda}}

\newcommand{\Sig}{{\Sigma}}

\newcommand{\al}{{\alpha}}
\newcommand{\bt}{{\beta}}

\newcommand{\de}{{\delta}}

\newcommand{\zt}{{\zeta}}
\newcommand{\te}{{\theta}}

\newcommand{\lm}{{\lambda}}

\newcommand{\vphi}{{\varphi}}

\newcommand{\bsk}{{\boldsymbol{k}}}

\newcommand{\bsq}{{\boldsymbol{q}}}

\newcommand{\bsx}{{\boldsymbol{x}}}
\newcommand{\bsy}{{\boldsymbol{y}}}

\newcommand{\bsE}{{\boldsymbol{E}}}
\newcommand{\bsG}{{\boldsymbol{G}}}
\newcommand{\bsK}{{\boldsymbol{K}}}

\newcommand{\bsX}{{\boldsymbol{X}}}
\newcommand{\bsnl}{{\boldsymbol{0}}}

\newcommand{\cD}{{\mathcal{D}}}
\newcommand{\cF}{{\mathcal{F}}}
\newcommand{\cG}{{\mathcal{G}}}
\newcommand{\cK}{{\mathcal{K}}}

\newcommand{\cO}{{\mathcal{O}}}

\newcommand{\lan}{{\langle}}

\newcommand{\nn}{{\nonumber}}

\newcommand{\pd}{{\partial}}

\newcommand{\ran}{{\rangle}}

\newcommand{\wt}{\widetilde}

\def\bbra{{\langle\kern-2.5pt\langle}}
\def\kket{{\rangle\kern-2.5pt\rangle}}

\begin{document}

\baselineskip=18pt

\begin{titlepage}
\rightline{KOBE-COSMO-16-11}
\rightline{MAD-TH-16-05}
\begin{center}
\vskip 1.5 cm
{\large \bf {Holographic non-Gaussianities in general single-field inflation}}\\[5mm]
\vskip 1cm
{
 {Hiroshi Isono,$^{1}$ Toshifumi Noumi,$^{2,3}$ Gary Shiu,$^{2,4}$ Sam S.C. Wong$^2$ and Siyi Zhou$^2$}}
\vskip 1.0cm
{\it$\!^1$ Department of Physics, Faculty of Science, Chulalongkorn University, Bangkok 10330, Thailand \\[1mm]
$\!^2$ Department of Physics and Jockey Club Institute for Advanced Study,\\
 Hong Kong University of Science and Technology, Hong Kong\\[1mm]
 $\!^3$ Department of Physics, Kobe University, Kobe 657-8501, Japan
 \\[1mm]
$\!^4$ Department of Physics, University of Wisconsin-Madison, Madison, WI 53706, USA}\\

\vskip 1.5cm

{\bf Abstract}
\end{center}

\noindent

We use holographic techniques to compute inflationary non-Gaussianities for general single-field inflation, including models with a non-trivial sound speed. In this holographic approach, the inflationary dynamics is captured by a relevant deformation of the dual conformal field theory (CFT) in the UV, while the inflationary correlators are  computed by conformal perturbation theory. 
 In this paper, we discuss the effects of higher derivative operators, such as $(\partial_\mu\phi\partial^\mu\phi)^{m}$, which  are known to induce a non-trivial sound speed and source potentially large non-Gaussianities. We compute the full   inflationary bispectra from the deformed CFT correlators. We also discuss the squeezed limit of the bispectra 
  from the viewpoint of operator product expansions. As is generic in the holographic description of  inflation, our power spectrum is blue tilted in the UV region. We extend our bispectrum computation to the IR region by resumming the conformal perturbations to all orders. We provide a self-consistent setup which reproduces a red tilted power spectrum, as well as all possible bispectrum shapes in the slow-roll regime.

\end{titlepage}

\setcounter{tocdepth}{2}
\tableofcontents

\newpage
\section{Introduction}
\setcounter{equation}{0}

By now, there is overwhelming evidence for the celebrated AdS/CFT correspondence  \cite{Maldacena:1997re}. Given its vast successes, it is natural to wonder if there is a de Sitter counterpart.
Despite the lack of an explicit string theory construction of de Sitter space, an analysis of the asymptotic symmetries along the lines of \cite{Maldacena:1997re} suggests 
 a similar duality between de Sitter space and conformal field theory \cite{Strominger:2001pn,Witten:2001kn}. The dS/CFT correspondence states that physics in $(d+1)$ dimensional de Sitter space ${\rm dS}_{d+1}$ is dual to some conformal field theory ${\rm CFT}_d$ in $d$ dimensional Euclidean space, 
of which the geometry is a constant time slice of the de Sitter spacetime. 
The isometry group of ${\rm dS}_{d+1}$ matches with the conformal group of Euclidean ${\rm CFT}_d$, which is $SO(d+1,1)$. This is the starting point of the 
 conjectured relationship between these two descriptions.

\medskip
While the dS/CFT correspondence was originally introduced to discuss quantum nature of gravity on de Sitter space, it has also shed some light on observational cosmology, particularly in the context of cosmic inflation. An inflationary universe may be regarded as a quasi-de Sitter space with an approximately (but not exactly) constant Hubble parameter $H$. Since the time-translation invariance of de Sitter space corresponds to the scale invariance of the dual CFT, the dS/CFT correspondence has to be extended to non-conformal theories in order to capture inflationary dynamics in a holographic manner. Just as the standard AdS/CFT correspondence, it was achieved, e.g. by performing relevant deformations in the dS/CFT correspondence~\cite{Strominger:2001gp,Larsen:2002et,Maldacena:2002vr,Larsen:2003pf,vanderSchaar:2003sz,Seery:2006tq,McFadden:2009fg,Anninos:2011ui,McFadden:2010vh,Bzowski:2011ab,Smolkin:2012er,Schalm:2012pi,Bzowski:2012ih,Mata:2012bx,Garriga:2013rpa,Anninos:2013rza,McFadden:2013ria,Pimentel:2013gza,Ghosh:2014kba,Garriga:2014ema,Garriga:2014fda,McFadden:2014nta,Kundu:2015xta}. This extended relationship is sometimes dubbed the inflation/deformed CFT correspondence.

\begin{figure}[t]\label{timeev}
  \centering
  \includegraphics[width=0.7\textwidth]{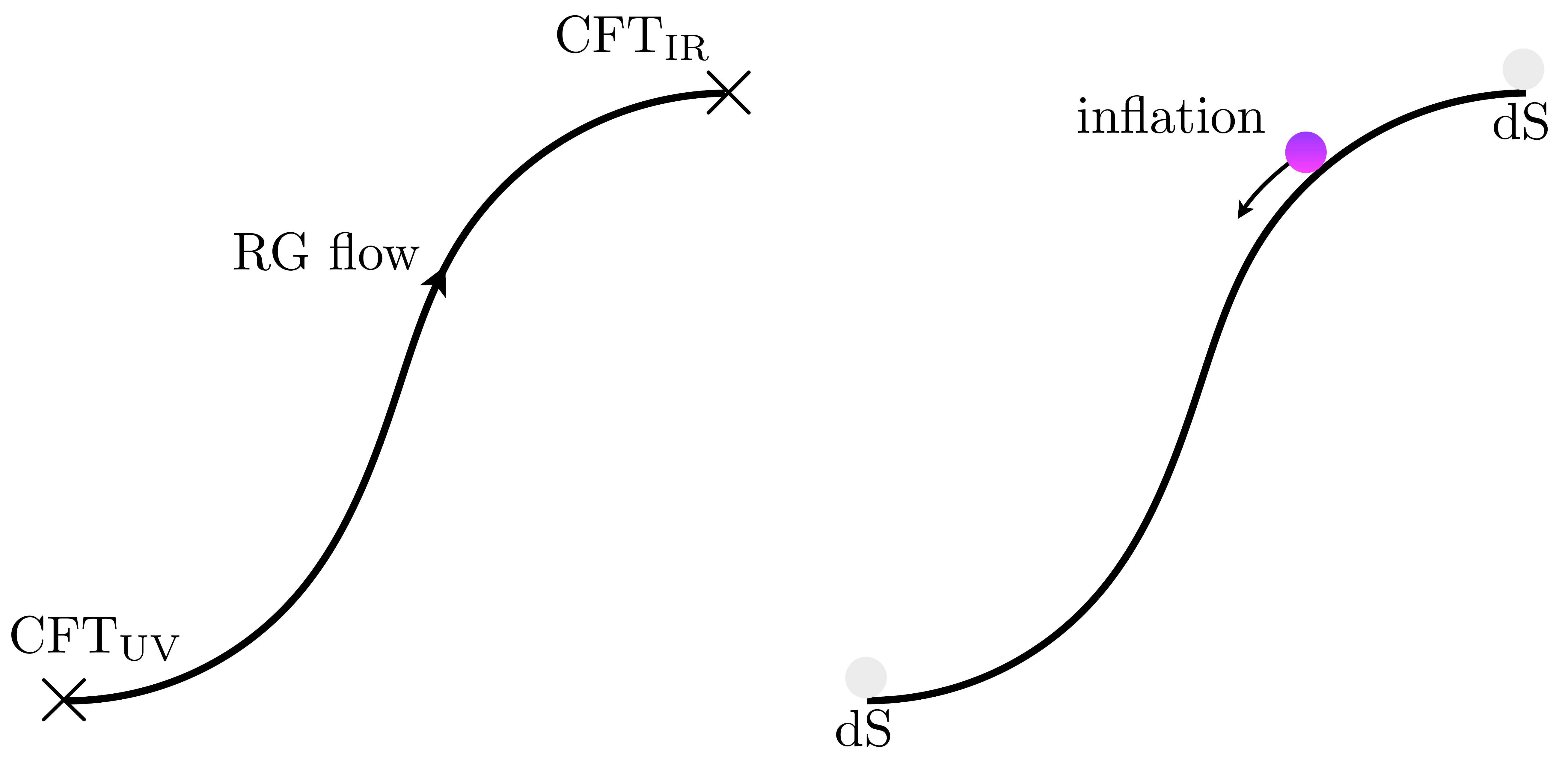}
  \caption{A sketch of the bulk and the dual CFT description of the cosmic evolution.
From the bulk perspective (right figure), the inflaton starts from the grey ball on the top of the potential and rolls all the way down to the bottom of the potential.
From the CFT perspective (left figure)
such an evolution is identified with an RG flow connecting two conformal fixed points.}
\end{figure}

\medskip
In this holographic picture the expansion history of our universe may be identified with 
the renormalization group (RG) flow of the field theory dual~\cite{Strominger:2001gp}. In particular the universe is asymptotically de Sitter space at the future and past infinities, corresponding to the UV and IR fixed points of the flow, with inflation in the interim (see Fig.\,\ref{timeev}). The slow-roll property of inflation is 
translated to the smallness of the beta function of the field theory dual.
Based on this picture, inflationary correlation functions have been computed for
some inflationary models via the relevant deformation of the field theory dual~\cite{vanderSchaar:2003sz,Schalm:2012pi,Garriga:2013rpa,McFadden:2013ria,Garriga:2014ema,Garriga:2014fda}. It was also used to derive inflationary consistency relations based on the broken conformal symmetry~\cite{Schalm:2012pi,Kundu:2015xta}.

\medskip
In light of 
these developments, we would like to use holographic techniques to compute inflationary non-Gaussianities for general single-field inflation \cite{Chen:2006nt}. Primordial non-Gaussianities, which 
give a direct measurement of interactions during inflation, are one of the most important probes of high energy physics in the early universe. While the level of non-Gaussianities generated by single-field inflationary models with a canonical kinetic term is suppressed by slow-roll parameters\cite{Maldacena:2002vr, Acquaviva:2002ud}, higher derivative interactions in the generalized Lagrangians are known to source potentially large non-Gaussianities as well as a non-trivial sound speed $c_s<1$~\cite{ArmendarizPicon:1999rj, Alishahiha:2004eh, Chen:2006nt}. The inflationary bispectrum for general single-field inflation has been computed directly on the inflation side long ago \cite{Chen:2006nt} (see \cite{Seery:2005wm} for a less complete result that applies only to cases with $c_s \approx 1$).
It was shown that the bispectrum is completely determined by five parameters: the usual slow-roll parameters $\epsilon, \eta$, the sound speed $c_s$ and its rate of change $s := \dot{c_s}/\left(c_s H\right)$, and $\Lambda$ which characterizes the derivative interactions \cite{Chen:2006nt}.\footnote{Our parameter $\Lambda$ was denoted by $\lambda$ in \cite{Chen:2006nt}. We use the capital letter $\Lm$ because $\lm$ is reserved for the anomalous conformal dimension of the deformation operator $O_0$, which is one of the key parameters appearing recurrently in this paper. See the next section for its definition.} 
In this paper, as a first step, we compute the bispectrum of general single-field inflation along the line of conformal perturbation theory of the CFT dual.

\medskip
In order to carry out the conformal perturbation theory, we need to specify the reference CFT. However, while a concrete realization of the dS/CFT correspondence was proposed for Vasiliev's higher spin gravity~\cite{Anninos:2011ui,Anninos:2013rza}, an explicit realization for Einstein gravity is yet unknown. The holographic studies of inflation in the literature are therefore mainly classified into the following three categories: 
\begin{enumerate}
\item Specify a concrete model of the reference CFT and the RG flow, and investigate the dual inflationary dynamics~\cite{Bzowski:2011ab, Smolkin:2012er}.
\item
Clarify model-independent properties of holographic inflation such as an expression of the power spectrum in terms of the beta function of the dual QFT~\cite{Schalm:2012pi, Bzowski:2012ih, Garriga:2013rpa, Garriga:2014ema}, and holographic derivation of consistency relations~\cite{Schalm:2012pi,Pimentel:2013gza,McFadden:2014nta}.  

\item
Specify the hypothetical reference CFT based on a concrete bulk model and the dS/CFT correspondence (sometimes called the holographic CFT in the context of AdS/CFT correspondence), and use conformal perturbation theory to compute inflationary spectra~\cite{Seery:2006tq,McFadden:2013ria, Ghosh:2014kba}. 
\end{enumerate}
As we will see, the shape of the inflationary bispectrum depends on details of the CFT dual. We therefore consider that the last approach is adequate to discuss how the bispectrum shape of general single-field inflation is realized through the conformal perturbation theory. Also, it will be a useful step for further understanding of primordial non-Gaussianities from the holographic viewpoint. Based on these motivations, we employ the third approach in the above to compute the bispectrum of general single-field inflation. In particular we reproduce the shape of the bispectrum dictated by the sound speed $c_s$, albeit in a perturbative manner order by order in $1-c_s$.

\medskip
The organization of the paper and main results of each section are as follows. In Sec.\,\ref{basicsaboutholography} we review basics of holographic inflation.
In particular we introduce the holographic dictionary between inflationary observables and correlators of the deformed CFT.

\medskip
In Sec.\,\ref{Sec:UV} we compute the inflationary bispectrum in a holographic way, using the conformal perturbation theory around the UV fixed point.
We first summarize general features of holographic primordial spectra which do not depend on details of inflationary models. In particular we derive consistency relations of bispectra based on the conformal Ward identity, by extending the previous derivation at the lowest order of the conformal perturbation~\cite{Schalm:2012pi}. We then compute the bispectrum of general single-field inflation at the UV scale.
Based on the third approach mentioned above,
we specify the UV CFT correlators by solving the Dirichlet boundary problem of the AdS model associated with the dS dynamics of our interests.
We also discuss why the bispectrum associated with higher derivative interactions vanishes in the squeezed limit,
from the viewpoint of operator product expansion,
which is a less model-sensitive argument based on conformal symmetry.

\medskip
The inflationary spectra around the UV fixed point are known to be blue tilted.
In Sec.\,\ref{sec:IR}, in order to reproduce a red tilted spectrum consistent with observation \cite{Ade:2015xua}, we extend our computation of the bispectrum to the IR region by taking into account the conformal perturbation at all orders. Under the approximation that the deformation operator is nearly marginal and the derivative interactions are reasonably small, we compute the leading contribution to the bispectrum at the IR scale and explicitly perform the resummation over the deformation parameter. We reproduce all possible bispectrum shapes of general single-field inflation in the slow-roll regime, even at the scale with the red spectrum beyond the UV scale. This is our main result in this paper. We relegate some useful details of CFT correlation functions to the appendices. 

\section{Holographic approach to single field inflation}\label{basicsaboutholography}
\setcounter{equation}{0}

In this section we summarize the basics of holographic inflation relevant to this paper \cite{Maldacena:2002vr,Schalm:2012pi,Garriga:2013rpa,McFadden:2013ria}. In Sec.\,\ref{21} we first sketch how the relevant deformation of the dual CFT can be related to the inflationary dynamics. We then in Sec.\,\ref{subsec:dictionary} introduce the relation between primordial spectra and correlation functions of the field theory dual. We will apply the general results introduced here to concrete inflationary models in the following sections.

\subsection{Inflation from relevant perturbations}\label{21}

In the holographic approach, the inflationary dynamics is captured by a relevant deformation of the dual CFT. We perturb this CFT by a relevant operator $ {O}_0$ dual to the inflaton field 
as
\begin{align}
\label{modified_action}
S[\chi]=S_{\rm CFT}[\chi]+\int d^3x\,\bar{\phi}\, {O}_0(\bsx)\,,
\end{align}
where $S_{\rm CFT}[\chi]$ is the action of the CFT, which we call the UV CFT, and $\chi$ stands for the integration variables (matter fields) of the path integral defining the UV CFT.
As depicted in Fig.\,\ref{timeev}, the UV CFT is associated with the future infinity of an asymptotically de Sitter space.
The deformation parameter $\bar\phi$ is related to the inflaton background value at late time corresponding to the Wilsonian cutoff scale $\Lambda$ of the field theory dual  as $\phi=\bar{\phi}\Lambda^{-\lambda}$.
We take the limit $\Lambda\to\infty$ with $\bar{\phi}$ being fixed.

\medskip
Just as the evolution of inflaton background deforms the de Sitter space and breaks some isometries, the relevant deformation $ {O}_0$ breaks the conformal invariance of the dual CFT. It is instructive to see this concretely from the scale dependence of correlation functions, with an analogy to the inflationary perturbations. Correlation functions in the deformed QFT can be expressed in terms of the UV CFT correlators as
\begin{align}
\langle\,\ldots\,\rangle=\left\langle\,\ldots\, \exp\left[-\int d^3x\,\bar{\phi}\, {O}_0(\bsx)\right]\,\right\rangle_{\rm CFT}\,,
\end{align}
where $\langle\,\ldots\,\rangle_{\rm CFT}$ is the correlation function in the UV CFT. In particular the two-point function in momentum space is given by\footnote{For a correlation function in momentum space we will put a prime on the bracket  $\lan...\ran'$, which means dropping the factor $(2\pi)^3\de^3(\sum\bsk_i)$, while for a correlation function in position space we will not put a prime on the ket.}
\begin{align}
\label{perturbedCFT}
\langle  O_0(\boldsymbol k) O_0(-\boldsymbol k) \rangle^\prime &= \langle  O_0 (\boldsymbol k)  O_0 (-\boldsymbol k) \rangle^\prime_{\operatorname{CFT}}-\bar \phi \langle  O_0 (\boldsymbol k)   O_0 (-\boldsymbol k)  O_0 (\boldsymbol 0)\rangle^\prime_{\operatorname{CFT}}
+\mathcal O(\bar{\phi}^2)\,.
\end{align}
We have also used $\int d^3x\, {O}_0(\bsx)= {O}_0({\boldsymbol k}=\boldsymbol 0)$. The scale dependence of each CFT correlator is determined by the dilatation symmetry as
\begin{align}
 \langle  O_0 (\boldsymbol k)   O_0 (-\boldsymbol k) \rangle^\prime_{\operatorname{CFT}}=A_0 k^{3-2\lambda}\,,
 \quad
 \langle  O_0 (\boldsymbol k)  O_0 (-\boldsymbol k)  O_0 (\boldsymbol 0)\rangle^\prime_{\operatorname{CFT}}=A_1 k^{3-3\lambda}\,,
 \label{A0A1}
\end{align}
where $\lambda$ is a positive constant defined by $\lambda=3-\Delta$ with $\Delta$ being the scaling dimension of $ {O}_0$ in the UV CFT. The coefficients $A_0$ and $A_1$ are determined only from the UV CFT data. We then have
\begin{align}\label{conformalperturbationtheory}
\langle   O_0 (\boldsymbol k)  O_0 (-\boldsymbol k) \rangle^\prime = A_0 k^{3-2\lambda}\left[1-\frac{A_1}{A_0}\bar \phi k^{-\lambda}+\mathcal O(\bar \phi^2)\right]\,.
\end{align}
Since the deformation parameter $\bar{\phi}$ always appears with a factor $k^{-\lambda}$, the correlation functions reduce to the UV CFT ones in the limit $k\to\infty$. On the other hand, the deformation becomes important in the IR as $k\to0$. The deformation induced momentum dependence probes the QFT at different scales,
as each mode of the observed inflationary perturbations probes the geometry at its own horizon crossing time.
Since holography converts the scale dependence of the QFT to the time evolution of the bulk geometry, we can compute inflationary correlation functions from the QFT correlators at an appropriate scale associated with the geometry of our interest  (see Fig.\,\ref{timeev}). Therefore one can extract inflationary observables from QFT correlators. For instance, the spectral index $(n_s-1)$ and its running are  in principle encoded in (\ref{conformalperturbationtheory}). 

\subsection{Holographic dictionary}\label{22}
\label{subsec:dictionary}

We next introduce the holographic dictionary translating the inflationary correlation functions and the QFT correlators. For this purpose, it is convenient to start from an action with sources,
\begin{align}
S[\chi;g_{ij},\phi]=S_{\rm CFT}[\chi;g_{ij}]+\int d^3x\sqrt{g}\phi {O}_0\,,
\end{align}
where $g_{ij}(\boldsymbol x)$ and $\phi(\bsx)=\bar{\phi}+\varphi(\boldsymbol x)$ source the energy-momentum tensor and the operator $ {O}_0(\boldsymbol x)$, respectively. The original action~\eqref{modified_action} is reproduced by setting $g_{ij}=\delta_{ij}$ and $\phi=\bar{\phi}$. The key idea of holography is that the bulk fields source operators in the dual QFT. For example, the scalar curvature perturbation $\zeta$ of inflation can be identified with the metric $g_{ij}$ as
\begin{align}
g_{ij}=e^{2\zeta}\delta_{ij}\,,
\end{align}
and sources the trace of the energy momentum tensor defined by
\begin{align} \label{Texp}
T& := g^{ij}\frac{2}{\sqrt{g}}\frac{\delta S}{\delta g^{ij}}
=-e^{-3\zeta}\frac{\delta S}{\delta \zeta}
=-\left(\frac{\partial\mathcal{L}}{\partial\zeta}+3\mathcal{L}\right)
\,,
\end{align}
where we introduced the Lagrangian density $\mathcal{L}$ such that $S=\int d^3x\sqrt{g}\mathcal{L}=\int d^3x\,e^{3\zeta}\mathcal{L}$. 
Here and in what follows we drop the tensor modes and focus on the scalar sector only. For later use, it is convenient to expand the energy-momentum tensor in $\zeta$ and $\varphi$ as
\begin{align}
\label{T_expand}
T&=T_0+T_1\zeta-3\varphi {O}_0+\ldots
\,,
\end{align}
where the dots stand for the second and higher orders in $\zeta$ and $\varphi$. In particular we may write $T_0$ and $T_1$ in terms of the Lagrangian density as
\begin{align}
T_0=-\left[\frac{\partial\mathcal{L}}{\partial\zeta}+3\mathcal{L}\right]_{\substack{\zeta=0\\\varphi=0}}\,,
\quad
T_1=-\left[\frac{\partial^2\mathcal{L}}{\partial\zeta^2}+3\frac{\partial\mathcal{L}}{\partial\zeta}\right]_{\substack{\zeta=0\\\varphi=0}}\,.
\end{align}
The action can then be expanded in $\zeta$ up to the second order as
\begin{align}
\nonumber
S[\chi;g_{ij},\phi]
&=S[\chi;\delta_{ij},\bar{\phi}]+\int d^3x\Bigg[\left[\frac{\partial\mathcal{L}}{\partial\zeta}+3\mathcal{L}\right]_{\substack{\zeta=0\\\varphi=0}}\,\zeta+\frac{1}{2}\left[\frac{\partial^2\mathcal{L}}{\partial\zeta^2}+6\frac{\partial\mathcal{L}}{\partial\zeta}+9\mathcal{L}\right]_{\substack{\zeta=0\\\varphi=0}}\,\zeta^2+\ldots\Bigg]
\\
\label{Sexpanded}
&=S[\chi;\delta_{ij},\bar{\phi}]-\int d^3x\left[T_0\zeta+\frac{1}{2}(3T_0+T_1)\zeta^2+\ldots\right]\,,
\end{align}
where note that the dots contain terms with $\varphi$ as well as higher order terms in $\zeta$. In the following we mostly use $U=3T_0+T_1$ instead of $T_1$ for notational simplicity.

\medskip
The holographic dictionary relates the bulk wave function $\Psi[\zeta]$ and the QFT partition function $Z[\zeta]$ as~\cite{Maldacena:2002vr}
\begin{align} \label{PsiZ}
\Psi[\zeta]=Z[\zeta]\,.
\end{align}
Here $\Psi[\zeta]$ is the wavefunction of $\zeta$ at the future infinity in the gauge $\phi(\boldsymbol x,t)=\bar{\phi}(t)$.\footnote{See, e.g., \cite{Mata:2012bx,Garriga:2013rpa,Ghosh:2014kba,Kundu:2015xta} for other gauge choices.}
On the other hand, we define the partition function in the dual QFT by
\begin{align}
Z[\zeta] := \int [d\chi] e^{-S[\chi;g_{ij},\bar{\phi}]}
\quad
{\rm with}
\quad g_{ij}=e^{2\zeta}\delta_{ij}\,.
\end{align}
We normalize it so that $Z[\zeta=0]=1$. It can then be expanded in $\zeta$ as
\begin{align}
\nonumber
Z[\zeta]&=
\exp\bigg[
\frac{1}{2}\int d^3x_1d^3x_2
\langle T_0(\bsx_1)T_0(\bsx_2)\rangle
\zeta(\bsx_1)\zeta(\bsx_2)
\nn\\
& \!\!\!\!
+\frac{1}{6}\int d^3x_1d^3x_2d^3x_3
\left[
\langle T_0(\bsx_1)T_0(\bsx_2)T_0(\bsx_3)\rangle
+3\langle T_0(\bsx_1)U(\bsx_2)\rangle\delta^{3}(\bsx_2-\bsx_3)
\right]\zeta(\bsx_1)\zeta(\bsx_2)\zeta(\bsx_3) \nn\\
& \!\!\!\! +\mathcal{O}(\zeta^4)\bigg]\,.
\end{align}
Here and in what follows we drop ultra-local terms, which contain two or more delta functions and are not important for our purpose. We also assume that one-point functions vanish.\footnote{If one takes into account the quadratic order $\frac{1}{2}T_2\zeta^2$ in $T$ in eq.~(\ref{T_expand}), it contributes to a contact term in the coefficient of $\zeta(\bsx)^3$, where all operators overlap at the same point $\bsx$. In principle, these contact terms are canceled by local counterterms.}  The correlation functions $\lan...\ran$ on the right hand side are defined with the action $S[\chi;\de_{ij},\bar\phi]$. By going into the momentum space and identifying $Z[\zeta]$ with the wave function $\Psi[\zeta]$, we can write primordial spectra in terms of correlation functions in the dual QFT as 
\begin{align}
\label{dictionaryTT}
\langle \zeta ({\boldsymbol k})\zeta(-{\boldsymbol k})\rangle^\prime &= \frac{1}{-2{\rm Re}\langle T_0 ({\boldsymbol k}) T_0 (-\boldsymbol k)  \rangle^\prime }\,,
\\
\label{dictionaryTTT}
\langle \zeta (\boldsymbol k_1) \zeta (\boldsymbol k_2) \zeta (\boldsymbol k_3) \rangle^\prime &= \frac{2{\rm Re}\langle T_0  (\boldsymbol k_1) T_0 (\boldsymbol k_2) T_0 (\boldsymbol k_3)   \rangle^\prime+2{\rm Re}\langle T_0(\boldsymbol k_1)U(-\boldsymbol k_1)\rangle^\prime +{\rm 2\, perms}}{\prod_{i=1}^3\left[-2{\rm Re}\langle T_0(\boldsymbol k_i)T_0(-\boldsymbol k_i) \rangle^\prime\right]}
\,,
\end{align}
where the correlators of $\zeta$ on the left hand side are defined by $\lan\, ... \,\ran=\int[d\zt]\,...\,|\Psi[\zt]|^2$.

\medskip
Finally, we reformulate the dictionary \eqref{dictionaryTT}-\eqref{dictionaryTTT} in terms of correlation functions of $ {O}_0$ by using the conformal Ward-Takahashi identity~\cite{Osborn:1989td,Baume:2014rla} 
\begin{equation}\label{master}
\langle T(\bsx) \rangle_s = -\lambda\phi(\bsx)\langle  O_0(\bsx)\rangle_s
= -\lambda\left(\bar\phi+\varphi(\bsx)\right)\langle   O_0(\bsx)\rangle_s\,,
\end{equation}
where $\langle\ldots\rangle_s$ is the correlation function computed in the presence of the sources $\zeta$ and $\varphi$. Expanding this master equation in $\zeta$ and $\varphi$, we can convert the $T_0$ correlators into $ {O}_0$ correlators. For example, the order $\zeta^1$ and $\varphi^1$ terms of Eq.\,\eqref{master} are given by 
\begin{align}
\langle T_0 ({\boldsymbol k}) T_0 (-\boldsymbol k)  \rangle^\prime+\langle T_1 (\boldsymbol 0) \rangle^\prime&=-\lambda\bar{\phi}\langle T_0 ({\boldsymbol k})  {O}_0(-\boldsymbol k)  \rangle^\prime\,,
\\
\langle T_0 ({\boldsymbol k})  {O}_0 (-\boldsymbol k)  \rangle^\prime{+}3\langle  {O}_0 (\boldsymbol 0) \rangle^\prime&=-\lambda\bar{\phi}\langle  {O}_0 ({\boldsymbol k})  {O}_0(-\boldsymbol k)  \rangle^\prime+\lambda\langle  {O}_0 (\boldsymbol 0) \rangle^\prime\,,
\end{align}
where we used the expression~\eqref{T_expand}. We then arrive at the relation,
\begin{align}
\langle T_0 ({\boldsymbol k}) T_0 (-\boldsymbol k)  \rangle^\prime
&=\lambda^2\bar{\phi}^2\langle  {O}_0 ({\boldsymbol k})  {O}_0(-\boldsymbol k)  \rangle^\prime
+(3-\lambda)\lambda\bar{\phi}\langle  {O}_0 (\boldsymbol 0) \rangle^\prime
-\langle T_1 (\boldsymbol 0) \rangle^\prime \nn\\
&=\lambda^2\bar{\phi}^2\langle  {O}_0 ({\boldsymbol k})  {O}_0(-\boldsymbol k)  \rangle^\prime\,,
\end{align}
where at the second equality we assumed that one-point functions of $ {O}_0$ and $T_1$ vanish. Similarly we can derive the relation, 
\begin{align}
\nonumber
&\langle T_0(\boldsymbol k_1)T_0(\boldsymbol k_2)T_0(\boldsymbol k_3) \rangle^\prime
+\langle U(\boldsymbol k_1)T_0(-\boldsymbol k_1)\rangle^\prime+\langle U(\boldsymbol k_2)T_0(-\boldsymbol k_2)\rangle^\prime+\langle U(\boldsymbol k_3)T_0(-\boldsymbol k_3)\rangle^\prime \\*
&= -\lambda^3\bar{\phi}^3\langle  O_0(\boldsymbol k_1)O_0(\boldsymbol k_2)O_0(\boldsymbol k_3) \rangle^\prime \nn\\
& \quad +\lambda^3\bar{\phi}^2 \Big[ \langle  O_0(\boldsymbol k_1)  O_0(-\boldsymbol k_1)\rangle^\prime  + \langle   O_0(\boldsymbol k_2)  O_0(-\boldsymbol k_2)\rangle^\prime  + \langle   O_0(\boldsymbol k_3) O_0(-\boldsymbol k_3)\rangle^\prime\Big]\,.
\end{align}
We therefore have
\begin{align}
\label{dictionaryOO}
\langle \zeta ({\boldsymbol k})\zeta(-{\boldsymbol k})\rangle^\prime &= \frac{1}{\lambda^2\bar{\phi}^2 \left[-2 {\rm Re}  \langle   O_0 ({\boldsymbol k})   O_0 (-\boldsymbol k)  \rangle^\prime \right]}\,,
\\
\label{dictionaryOOO}
\langle \zeta (\boldsymbol k_1) \zeta (\boldsymbol k_2) \zeta (\boldsymbol k_3) \rangle^\prime &= \frac{-2\bar{\phi} {\rm Re}\langle   O_0 (\boldsymbol k_1)   O_0 (\boldsymbol k_2)  O_0 (\boldsymbol k_3)   \rangle^\prime+2{\rm Re}\langle  O_0(\boldsymbol k_1)   O_0(-\boldsymbol k_1)\rangle^\prime +{\rm 2\, perms}}{\lambda^{{3}}\bar{\phi}^{{4}}\prod_{i=1}^3\left[-2{\rm Re}\langle  O_0(\boldsymbol k_i) O_0(-\boldsymbol k_i) \rangle^\prime\right]}\,.
\end{align}
Interestingly, no correlation functions including $U$ appear in these relations. As a result we can compute primordial spectra purely in terms of correlation functions of ${O}_0$ together with the parameters $\lambda$ and $\bar\phi$.

\section{The UV story}
\label{Sec:UV}
\setcounter{equation}{0}

In this section we compute the inflationary bispectrum in a class of $P(X,\phi)$ inflation models by using the perturbed CFT presented in the last section. As we discussed in the last section, the perturbation from the UV CFT corresponds to the perturbative expansion in $\bar{\phi}k^{-\lambda}$. In this section we focus on the UV region, where higher order terms in $\bar{\phi}k^{-\lambda}$ are negligible. In the next section we extend our result to the IR region by performing a resummation.

\subsection{Generality and consistency relation of bispectrum}

Let us first discuss general properties of holographic inflationary correlation functions. As we mentioned earlier, correlation functions in the dual QFT are expanded in the deformation parameter $\bar{\phi}$. Each order of a two-point function in $\bar{\phi}$ is determined by the dilatation symmetry up to a constant as
\begin{align}
\langle O_0 ({\boldsymbol k}) O_0 (-\boldsymbol k)  \rangle^\prime
=k^{3-2\lambda}\sum_{n=0}^\infty A_n\left(-\bar\phi k^{-\lambda}\right)^n\,.
\end{align}
where $A_0$ and $A_1$ are given by \eqref{A0A1} and other $A_n$'s are defined in a similar way. They are uniquely determined by the UV CFT data, i.e., the spectrum and the OPE coefficients of the UV CFT. Note also that $A_n$'s are real constants because ${O}_0$ is a real scalar. Through the holographic dictionary \eqref{dictionaryOO}, the inflationary power spectrum is given by 
\begin{align}
\mathcal{P}_\zeta(k)=\frac{k^3}{2\pi^2}\langle \zeta ({\boldsymbol k})\zeta(-{\boldsymbol k})\rangle^\prime
=\frac{k^{2\lambda}}{2\pi^2(-2\lambda^2\bar{\phi}^2)}\left[\sum_{n=0}^\infty A_n\left(-\bar\phi k^{-\lambda}\right)^n\right]^{-1}\,.
\end{align}
We may compute the corresponding spectral index as
\begin{align}
\label{spectral_index}
n_s-1=\frac{d\ln \mathcal{P}_\zeta}{d \ln k}
&=2\lambda+\lambda\frac{\sum nA_n\left(-\bar\phi k^{-\lambda}\right)^n}{\sum A_n\left(-\bar\phi k^{-\lambda}\right)^n} \nn\\
&=\lambda\left[2+\frac{A_1}{A_0}(-\bar{\phi}k^{-\lambda})+\left(2\frac{A_2}{A_0}-\frac{A_1^2}{A_0^2}\right)\left(-\bar{\phi}k^{-\lambda}\right)^2+\ldots\right]\,.
\end{align}
Notice that the spectrum is blue tilted, i.e., $n_s-1>0$, in the UV region because the UV CFT is deformed by a relevant operator $\lm>0$. Therefore we need to go to the IR region to reproduce a red spectrum.

\medskip
In the inflationary context it is common to normalize the bispectrum by a factor of the power spectrum squared. By using the dictionary~\eqref{dictionaryOO} and~\eqref{dictionaryOOO}, we may express the normalized bispectrum
in terms of the $O_0$ correlators as
\begin{align}
\frac{\langle \zeta(\boldsymbol k_1)\zeta(\boldsymbol k_2)\zeta(\boldsymbol k_3)\rangle'}{\langle \zeta(\boldsymbol k_1)\zeta(-\boldsymbol k_1)\rangle'\langle \zeta(\boldsymbol k_2)\zeta(-\boldsymbol k_2)\rangle'}
&=-\lambda\left(1
+\frac{{\rm Re}\langle  O_0 ({\boldsymbol k}_2)   O_0 (-{\boldsymbol k}_2)  \rangle^\prime}{{\rm Re}\langle   O_0 ({\boldsymbol k}_3) O_0 (-{\boldsymbol k}_3)  \rangle^\prime}
+\frac{{\rm Re}\langle  O_0 ({\boldsymbol k}_1)  O_0 (-{\boldsymbol k}_1)  \rangle^\prime}{{\rm Re}\langle   O_0 ({\boldsymbol k}_3)   O_0 (-{\boldsymbol k}_3)  \rangle^\prime}\right) \nn\\
&\quad~
+\lambda\bar\phi\,\frac{{\rm Re}\langle  O_0 ({\boldsymbol k}_1)  O_0 ({\boldsymbol k}_2)   O_0 ({\boldsymbol k}_3)  \rangle^\prime}{{\rm Re}\langle   O_0 ({\boldsymbol k}_3)   O_0 (-{\boldsymbol k}_3)  \rangle^\prime}\,.
\label{shape_function}
\end{align}
By a similar argument for the power spectrum case, we may rewrite it in terms of the UV CFT correlators. Later we will do that explicitly for a concrete UV CFT, but in the rest of this subsection we focus on the squeezed limit without specifying the detail of the UV CFT. In the squeezed limit $k_1\ll k_2=k_3$, the normalized bispectrum~\eqref{shape_function} is reduced to the form,
\begin{align}
\label{squeezed_shape_function}
&\frac{\langle \zeta(\boldsymbol k_1)\zeta(\boldsymbol k_2)\zeta(\boldsymbol k_3)\rangle'}{\langle \zeta(\boldsymbol k_1)\zeta(-\boldsymbol k_1)\rangle'\langle \zeta(\boldsymbol k_2)\zeta(-\boldsymbol k_2)\rangle'}
\to-2\lambda+\lambda\bar\phi\,\frac{{\rm Re}\langle   O_0 (\bsnl) O_0 ({\boldsymbol k}_2)  O_0 (-{\boldsymbol k}_2)  \rangle^\prime}{{\rm Re}\langle  O_0 ({\boldsymbol k}_2)  O_0 (-{\boldsymbol k}_2)  \rangle^\prime}\,.
\end{align}
In the inflationary context, the consistency relation, i.e., the Ward-Takahashi identity for the broken de Sitter symmetry, tells us that the shape function in the squeezed limit is related to the spectral index $n_s$ as~\cite{Maldacena:2002vr,Creminelli:2004yq}
\begin{align}
\label{consistency_relation}
\frac{\langle \zeta(\boldsymbol k_1)\zeta(\boldsymbol k_2)\zeta(\boldsymbol k_3)\rangle'}{\langle \zeta(\boldsymbol k_1)\zeta(-\boldsymbol k_1)\rangle'\langle \zeta(\boldsymbol k_2)\zeta(-\boldsymbol k_2)\rangle'}\to -(n_s-1)\,,
\end{align}
where the spectral index is computed at the horizon crossing time of the modes ${\boldsymbol k_2=-\boldsymbol k_3}$. 

Let us  reproduce this relation from the QFT side using the expression \eqref{squeezed_shape_function}. Since the operator $ {O}_0(\bsnl)$ with zero momentum is sourced by the deformation parameter $\bar{\phi}$, as is seen from the deformed action~\eqref{modified_action}, we may rewrite the second term of~\eqref{squeezed_shape_function} as
\begin{align}
\lambda\bar{\phi}\frac{{\rm Re}\langle  O_0 (\bsnl) O_0 ({\boldsymbol k}_2)  O_0 (-{\boldsymbol k}_2)  \rangle^\prime}{{\rm Re}\langle  O_0 ({\boldsymbol k}_2)   O_0 (-{\boldsymbol k}_2)  \rangle^\prime}
=-\lambda\bar\phi\frac{\partial_{\bar{\phi}}{\rm Re}\langle  O_0 ({\boldsymbol k}_2) O_0 (-{\boldsymbol k}_2)  \rangle^\prime}{{\rm Re}\langle O_0 ({\boldsymbol k}_2)  O_0 (-{\boldsymbol k}_2)  \rangle^\prime}
=-\lambda\frac{\sum nA_n\left(-\bar\phi k_2^{-\lambda}\right)^n}{\sum A_n\left(-\bar\phi k_2^{-\lambda}\right)^n}\,.
\end{align}
By comparing this with the spectral index in Eq.\,\eqref{spectral_index} {(after the replacement $k\to k_2$)}, we can readily find that the consistency relation \eqref{consistency_relation} holds for any inflationary three-point function computed via holography for any single-field inflation.\footnote{A similar derivation is given in~\cite{Schalm:2012pi} at the leading order in $\bar{\phi}$. Our derivation can be thought of as its all-order extension. Another type of holographic derivation may be found in~\cite{Berezhiani:2013ewa,McFadden:2014nta}, where an infinite set of consistency relations~\cite{Hinterbichler:2013dpa} were derived (at all orders in $\bar{\phi}$) based on diffeomorphism invariance of the field theory dual or equivalently  spatial diffeomorphism invariance on the constant time slice.
} 
It should be noticed that the dilatation symmetry of the UV CFT plays a crucial role in our derivation. As mentioned in the last section, the deformation parameter $\bar{\phi}$ appears in the two-point function always in the form $\bar{\phi}k^{-\lambda}$ as a consequence of the dilatation symmetry. As a result, we may convert a derivative in $\bar{\phi}$ into that in the momentum $k$,
\begin{align}
-\lambda\frac{\partial }{\partial \ln\bar{\phi}}\ln\left[ \sum_{n=1}^\infty A_n\left(-\bar\phi k^{-\lambda}\right)^n\right]
=\frac{\partial }{\partial \ln k}\ln\left[ \sum_{n=1}^\infty A_n\left(-\bar\phi k^{-\lambda}\right)^n\right]\,,
\end{align}
to reproduce the consistency relation.

\begin{figure}[t] 
  \centering
  \includegraphics[width=0.4\textwidth]{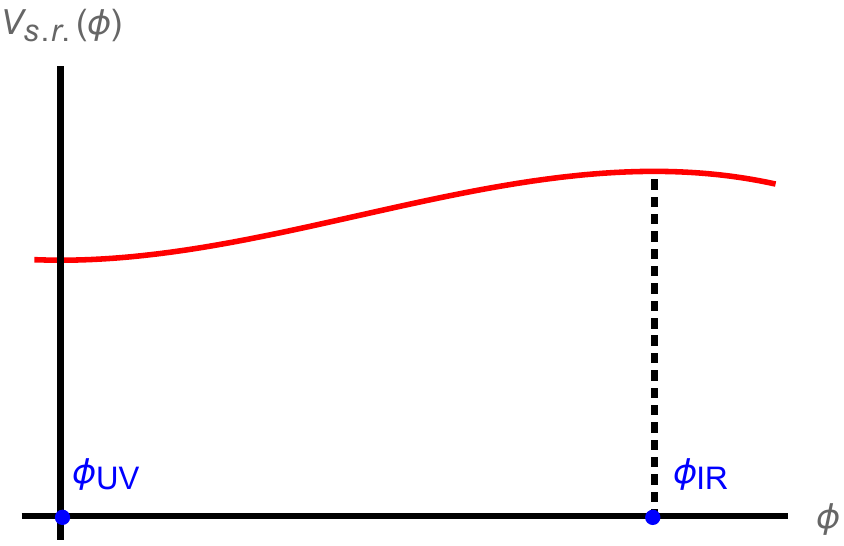}
  \caption{\label{fig:shape} Illustration of the slow-roll potential $V_{\rm s.r.}(\phi)$. }
\end{figure}

\subsection{Inflationary setup} \label{infsetup}

In the last subsection we showed that the three-point function in the squeezed limit is governed by the (broken) dilatation symmetry and its behavior does not depend on the details of the model. In contrast, the shape of three-point functions for general momentum configurations highly depends on the inflationary model. In the rest of this paper we specify a concrete UV CFT model and compute the shape function through the dictionary \eqref{dictionaryOOO} and the conformal perturbation theory.

\medskip
As we mentioned in Introduction, we would like to take into account higher derivative operators of general single-field inflation in the holographic inflation. To clarify our setup, let us begin by briefly reviewing the slow-roll inflationary models discussed in the holographic context:
\begin{align}
\nonumber
& S_{\rm s.r.} = \int d^4 x \sqrt{-g}\left[\frac{M_{\rm Pl}^2}{2}R -\frac{1}{2} (\partial_\mu\phi)^2 -V_{\rm s.r.}(\phi)\right] \\
\mbox{with ~ }
\quad
& V_{\rm s.r.}(\phi)=3M_{\rm Pl}^2H_0^2+\frac{m^2}{2} \phi^2 -\frac{g_3}{3}\phi^3 +\mathcal O(\phi^4)\,. \label{srpot}
\end{align}
Here the inflaton mass $m$ and the scaling dimension $\Delta=3-\lambda$ of the operator ${O}_0$ dual to the inflaton are related by $m^2=\lambda(3-\lambda)H_0^2$. We assume that $\lambda\ll1$ for the slow-roll property. We also assume that $g_3\sim 1$ to make the conformal perturbation work well throughout the RG flow generated by the deformation of the CFT.\footnote{More precisely, we assume that the mass term and the cubic interaction dominate over the other terms throughout the flow.} As depicted in Fig.\,\ref{fig:shape}, this potential has two extrema,
\begin{align} \label{extrema}
\phi_{\rm UV}=0\,,
\quad
\phi_{\rm IR}=\frac{3\lambda {H_0^2}}{g_3}+\mathcal{O}(\lambda^2)\,,
\end{align}
which correspond to the dual UV and IR CFTs, respectively.
We may find a time-dependent solution connecting the two extrema,
\begin{align}
\label{inflaton_flow}
\phi(t)=\phi_* \frac{1}{e^{\lambda H_0 (t-t_*)}+(1-e^{\lambda H_0 (t-t_*)})\frac{g_3\phi_*}{3\lambda {H_0^2}}}\,,
\end{align}
which satisfies the Hamilton-Jacobi equation for a spatially homogeneous configuration,
\begin{align}
\dot{\phi}\simeq-\lambda H_0\phi+\frac{g_3}{3H_0}\phi^2\,,
\end{align}
in the regime $\lambda\ll1$.
Here we introduced $\phi(t_*) = \phi_*$ with $t_*$ being some reference time.
The inflaton background evolution~\eqref{inflaton_flow} can then be identified with the RG flow connecting the two CFTs.
Also let us recall that the scalar spectral index is directly related to the scaling dimension of the deformation operator [see, e.g., Eq.~\eqref{spectral_index}]. 
Since the deformation operator is relevant (irrelevant) near the UV (IR) fixed point, the scalar power spectrum is blue (red) tilted in the UV (IR) region~\cite{Bzowski:2012ih,McFadden:2013ria,Garriga:2014ema}.

\medskip
{Let us now add higher derivative operators to the above slow-roll model. The general single-field inflation, or the so-called $P(X,\phi)$ model, has an action of the form,
\begin{align}
S = \int d^4 x \sqrt{-g}\left[\frac{M_{\rm Pl}^2}{2}R +P(X,\phi)\right]
\,,
\end{align}
where $P(X,\phi)$ is an arbitrary function of the inflaton $\phi$ and its kinetic operator $X=-\frac{1}{2}(\pd_\mu\phi)^2$. To discuss the bispectrum associated with higher derivative operators in general single-field inflation, it will be reasonable to focus on the following class of $P(X,\phi)$ model:\footnote{
In other words, we assume $P_{X\phi}=0$ (for any $\phi$ and $X$) for simplicity. Here and in what follows we use the notation such as $P_{X\phi}=\pd^2P/\pd X\pd\phi$.}
\begin{align}
\label{K_setup}
P(X,\phi)=X+\sum_{m \geq  2}\alpha_mX^{m}-V_{\rm s.r.}(\phi)\,,
\end{align}
where $V_{\rm s.r.}(\phi)$ is the slow-roll potential in Eq.~\eqref{srpot}.} Although our model \eqref{K_setup} accommodates stationary classical solutions $\phi=0$ and $\phi=\phi_{\rm IR}$, it does not necessarily mean that there exists a classical solution connecting the two extrema (corresponding to the RG flow connecting the two CFTs). Indeed, there is no such flow when $\alpha_m$'s are big, essentially because the slow-roll potential makes a subdominant contribution to the inflaton dynamics. In order to discuss a self-contained model with both the red tilted power spectrum and the non-Gaussianities in general single-field inflation, we focus on the following parameter region in this paper: for each $m \geq 2$
\begin{align}
\label{condition_alpha}
{\lambda\ll \alpha_m\lambda^{4(m-1)}\ll1\,. }
\end{align}
As we discuss in Sec.\,\ref{sec:IR}, there exists an RG flow connecting the two CFTs in this parameter region. Also the sound speed $c_s$ of the scalar perturbation satisfies $\lambda\ll c_s^{-2}-1\lesssim1$ in the IR region. As a result, the bispectrum associated with the non-trivial sound speed dominates over the slow-roll type one. This is the inflationary setup we consider in this paper. It should be noted that if we are not interested in reproducing a red tilted power spectrum, we do not have to introduce the slow-roll potential. We may also relax the condition~\eqref{condition_alpha} to realize a small sound speed $c_s\ll1$ and a large non-Gaussianity $f_{\rm NL}\gg1$ even in the UV region.
However, we focus on our parameter set to reproduce the red tilted spectrum in this paper.

\subsection{UV CFT} \label{UVCFT}

{
We then would like to specify the UV CFT corresponding to the local minimum $\phi=0$ to apply the conformal perturbation theory. As we mentioned in Introduction, we use the dS/CFT correspondence to specify the hypothetical UV CFT in this paper.

\medskip
As explained in~\cite{Maldacena:2002vr}, the dS/CFT correspondence can be related to the AdS/CFT correspondence by  an analytic continuation. If we write the Poincar\'e metrics of dS and Euclidean AdS as
\begin{align}
ds_{\rm dS}^2=R_{\rm dS}^2\eta^{-2}(-d\eta^2+d\bsx^2)\,,
\quad
ds_{\rm AdS}^2=R_{\rm AdS}^2z^{-2}(dz^2+d\bsx^2)\,,
\end{align}
they are related to each other by the analytic continuation\footnote{
The dS/CFT correspondence and AdS/CFT correspondence can also be related by an analytic continuation of the Planck mass~\cite{McFadden:2009fg}. The relation of these two approaches was discussed, e.g., in \cite{Garriga:2014fda}. }
\begin{align}
\label{dS_to_AdS}
z=-i\eta\,,\quad
R_{\rm AdS}=-iR_{\rm dS}\,,
\end{align}
where $R_{\rm dS}$ and $R_{\rm AdS}$ are the dS radius and the AdS radius, respectively. In particular, the dS radius is the inverse of the Hubble parameter~$H_0$.\footnote{The Hubble parameter of the exact dS (dual to the UV CFT) is given by $H_0$ introduced in Eq.~\eqref{srpot}, whereas the Hubble parameter $H(t)$ during inflation is time-dependent and deviates from $H_0$ because of the inflaton background.} Furthermore, we can show that the analytic continuation~\eqref{dS_to_AdS} relates the perturbation theory around $\phi=0$ (corresponding to an exact dS background) to the AdS perturbation theory of the following action:
\begin{align} \label{AdSaction}
S_{\rm AdS}=\int d^4x\sqrt{G} \bigg[
\frac{1}{2}(\pd\Phi)^2+V(\Phi)
+\sum_{m \geq 2}\frac{a_m}{2^m}(\pd\Phi)^{2m}
\bigg]
\end{align}
with the potential $V(\Phi)$ and the coupling $a_m$ of the form,
\begin{align}
V(\Phi)=\frac{m^2}{2}\Phi^2-\frac{g_3}{3}\Phi^3+\mathcal{O}(\Phi^4)\,,
\quad
a_m=(-1)^{m+1}\alpha_m\,.
\end{align}
Here for simplicity we neglected gravitational fluctuations, which are not relevant to our computation. We also introduced $G$ and $\Phi$ to denote the AdS metric and the scalar field, respectively, to avoid notational confusion. Just as the original argument in~\cite{Maldacena:2002vr}, our UV CFT correlation functions are then computed via the Witten diagrams for this AdS model followed by the replacement
\begin{align} \label{Mflip}
R_{\rm AdS}=-iR_{\rm dS}, \quad a_m=(-)^{m+1}\alpha_m\,.
\end{align}
In the following we will compute the hypothetical UV CFT correlation functions based on this approach. Since the calculation is a bit complicated essentially because of the singularity in the limit $\lambda\to0$, we put the details in appendices \ref{wittendiagram} and \ref{explicit}, and summarize the results and the essence of our computation in the main text. Also for simplicity we mostly set $R_{\rm dS}=1$ in the following.
}

\subsection{Bispectrum at the UV}
\label{subsec:Witten_diagram}
We then compute the three-point function of $O_0$ using the conformal perturbation theory for the model \eqref{K_setup}. As we mentioned earlier, in this section we focus on the UV region defined by
\begin{align} \label{uvregion}
\bar{\phi}k^{-\lambda}\ll \lambda\,,
\end{align}
where $k$ is a typical momentum scale of correlation functions of our interest. In this UV region, there are two types of the leading contributions to the three-point functions~\eqref{shape_function}. The first contribution is from the first term of~\eqref{shape_function} and is the zeroth order in $\bar{\phi}$:
\begin{align}\label{first_contribution}
-\lambda\left[
1+\left(\frac{k_2}{k_3}\right)^{3-2\lambda}+\left(\frac{k_1}{k_3}\right)^{3-2\lambda}
\right]\,,
\end{align}
which corresponds to the slow-roll type shape associated with the $\eta$ parameter~\cite{Maldacena:2002vr}.

\begin{figure}[t]
  \centering
  \includegraphics[width=0.3\textwidth]{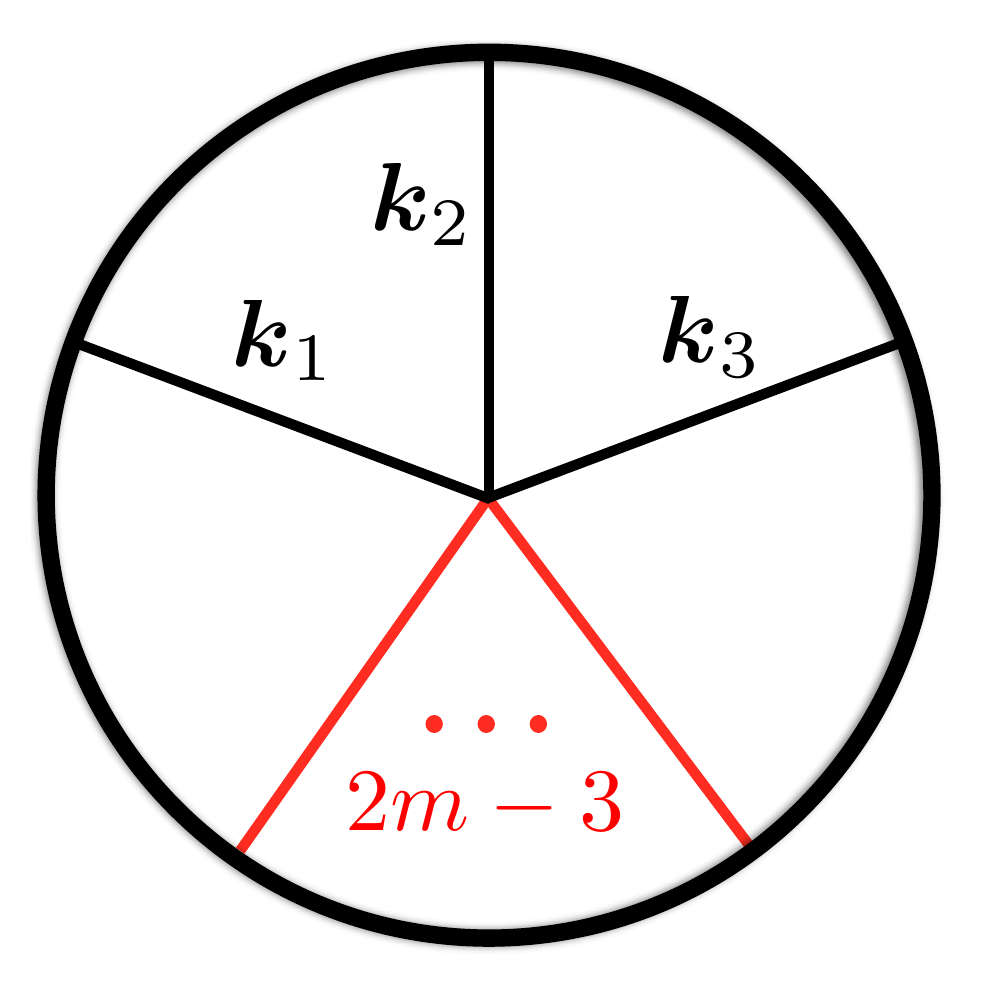}
  \caption{\small The Witten diagram for $\langle  O_0 ({\boldsymbol k}_1)  O_0 ({\boldsymbol k}_2)  O_0 ({\boldsymbol k}_3) O^{2m-3}_0 ({\boldsymbol 0})  \rangle^\prime_{\rm CFT}$. The black line denotes the bulk-boundary propagator $\mathcal K_{\boldsymbol k_i}(z)$. The red line represents the zero-momentum bulk-boundary propagator $\mathcal K_{\boldsymbol 0}(z)$.
  }\label{WittenDiagramaaa}
\end{figure}

\medskip
The other contribution is the $\alpha_m^1$-terms of the second term of~\eqref{shape_function}, which corresponds to the Witten diagram in Fig.\,\ref{WittenDiagramaaa}. 
In this paper, we are interested in the leading order in $\lm$ according to the parameter regime \eqref{condition_alpha}. Noting that the integral for this Witten diagram is not singular as $\lm\to 0$,\footnote{See also the comment at the end of Sec.\,\ref{alphamthree-point}.} we can easily evaluate the integral at the leading order in $\lm$. The result is 
\begin{align}
&\lambda\bar\phi\,\frac{{\rm Re}\langle  O_0 ({\boldsymbol k}_1)  O_0 ({\boldsymbol k}_2)  O_0 ({\boldsymbol k}_3)  \rangle^\prime}{{\rm Re}\langle O_0 ({\boldsymbol k}_3)  O_0 (-{\boldsymbol k}_3)  \rangle^\prime}
\ni \frac{4 k_1 k_2 k_3}{k_2^3} \mathcal F\left(\frac{k_1}{k_3},\frac{k_2}{k_3}\right)\,,
\label{second_contribution}
\end{align}
where the shape function $\cF$ is defined as 
\begin{align}
\mathcal F\left(\frac{k_1}{k_3},\frac{k_2}{k_3}\right) 
&=
\left[-\frac{4}{3}X^2P_{XXX}\frac{3k_1k_2k_3}{2(k_1+k_2+k_3)^3} 
+2XP_{XX}G(k_1,k_2,k_3)\right]_{X=(-\lm\bar\phi)^2/2}.
\end{align}
Here we introduced $P_{XX}=\pd^2P/\pd X^2$ and $P_{XXX}=\pd^3P/\pd X^3$. The function $G$ is defined as
\begin{align}
& G(k_1,k_2,k_3) \nn\\
\label{def_G}
&=
-\frac{k_1^2 k_2^2+k_1^2 k_3^2 + k_2^2 k_3^2}{k_1 k_2 k_3 ( k_1+k_2 +k_3)} 
+ \frac{k_1^2 k_2^3 +k_1^2 k_3^3 +k_2^2 k_3^3 +k_2^2 k_1^3 +k_3^2 k_1^3 +k_3^2 k_2^3}{2 k_1 k_2 k_3 (k_1+k_2+k_3)^2} 
+\frac{k_1^3 +k_2^3 +k_3^3}{8 k_1 k_2 k_3}.
\end{align}
This result reproduces the shapes associated with the $X^m$ coupling computed in \cite{Chen:2006nt}. We may observe that the first contribution~\eqref{first_contribution} is responsible for the consistency relation, whereas the second contribution \eqref{second_contribution} vanishes in the squeezed limit. Indeed, the inflationary parameters in the UV region are given by\footnote{
Our convention for the slow-roll parameters is $\epsilon= -\dot{H}/H^2$ and $\eta=\dot{\epsilon}/(\epsilon H)$, where $H(t)$ is the Hubble parameter.}
\begin{align}
\epsilon &= \frac{1}{2}\left(\lambda\bar{\phi}k^{-\lambda}\right)^2\,,  \\
\quad
\eta &= -2\lambda\,,  \\
\frac{1}{c_s^2}-1 &= \sum_{m \geq 2} \frac{m(m-1)}{2^{m-2}} \alpha_m \left(\lambda\bar{\phi}k^{-\lambda}\right)^{2m-2}\,, 
\\
s := \frac{\dot{c}_s}{Hc_s} &= \sum_{m \geq 2} \frac{m(m-1)^2}{2^{m-2}} {\lambda\, \alpha_m (\bar\lambda\phi k^{-\lambda})^{2m-2}}\,.
\end{align}
We may easily check that our three-point functions are consistent with the inflationary results in~\cite{Chen:2006nt}.

\medskip
In the rest of this section, we revisit the shape of bispectrum associated with the higher derivative operators, i.e., the $\mathcal{O}(\alpha_m)$ contribution \eqref{second_contribution} from the OPE perspective. In particular we discuss why it vanishes in the squeezed limit. Let us consider the CFT correlation functions of the form,
\begin{align}
\langle  O_0(\boldsymbol k_1)  O_0(\boldsymbol k_2) O_0(\boldsymbol k_3) O_0(\bsnl)\ldots O_0(\bsnl)\rangle'_{\rm CFT}\,.
\end{align}
As we have discussed, such correlation functions constitute three-point functions of the deformed CFT. In particular, they are relevant to the $\mathcal{O}(\alpha_m)$ contribution to the bispectrum. In the momentum space, the conformal partial wave expansion simply reads
\begin{align}
&\langle  O_0(\boldsymbol k_1) O_0(\boldsymbol k_2) O_0(\boldsymbol k_3) O_0(\bsnl)\ldots O_0(\bsnl)\rangle' _{\rm CFT} \nn\\[+5pt]
& \quad = 
\sum_I \frac{\langle O_0(\boldsymbol k_1)  O_0(\boldsymbol k_2) O_I(\boldsymbol k_3)\rangle'_{\rm CFT}\langle  O_I(\boldsymbol k_3)   O_0(-\boldsymbol k_3)  O_0(\bsnl)\ldots O_0(\bsnl) \rangle'_{\rm CFT}}{\langle  O_I(\boldsymbol k_3)  O_I(-\boldsymbol k_3) \rangle'_{\rm CFT}} \,,
\end{align}
where $ O_I$'s denote all the primary operators in the CFT. The shape of each term in the summation is determined by the three-point functions, $\langle  O_0(\boldsymbol k_1)  O_0(\boldsymbol k_2)  O_I(\boldsymbol k_3)\rangle'_{\rm CFT}$, which is uniquely fixed by the conformal symmetry up to an overall OPE coefficient. For example, when $O_I$ is a primary scalar, it takes the form,
\begin{align}
\nonumber
& \langle  O_{0} (\boldsymbol k_1) O_{0} (\boldsymbol k_2)  O_{I} (\boldsymbol k_3) \rangle'_{\rm CFT} \nn\\[+5pt]
&\quad
=C_{O_0O_0O_I} k_1^{\frac{3}{2}-\lambda}k_2^{\frac{3}{2}-\lambda}k_3^{\Delta_I-\frac{3}{2}} \int_0^\infty dx x^{\frac{1}{2}} K_{\frac{3}{2}-\lambda}(k_1 x)K_{\frac{3}{2}-\lambda}(k_2 x)K_{\Delta_I-\frac{3}{2}}(k_3 x)\,,
\end{align}
where $\Delta_I$ is the conformal dimension of $ {O}_I$. In Appendix~\ref{tripleKintegrals}, we compute CFT three-point functions with two nearly marginal primary scalars and one primary scalar with a conformal dimension close to a non-negative integer $n$. There we find that such three-point functions vanish in the squeezed limit $k_1\ll k_2=k_3$ except for the case $n=0,3$. In particular the bispectrum associated with derivative couplings is independent of the inflaton cubic coupling $g_3$, so that it will be dominated by the partial wave where $ {O}_I$ is composite (multi-trace), e.g., operators schematically of the form $ {O}_0\partial^p{O}_0$ ($p=0,1,2,\ldots$). Since $ {O}_0$ is nearly marginal, such composite operators have nearly integer dimensions with $n\geq6$, and therefore their partial wave contributions vanish in the squeezed limit. The bispectrum associated with the higher derivative operators $X^m$ then vanishes in the squeezed limit as long as this feature survives after summing up all the partial waves. 
Further progress in obtaining the precise momentum dependence would require a clarification of the OPE coefficients, which we postpone for future work. 

\section{Beyond the UV}
\label{sec:IR}
As described in Introduction, the bulk geometry of our setup is asymptotically de Sitter space both at the early and late time. In the dual QFT point of view, we have an RG flow between an IR CFT and a UV CFT. In~\cite{Bzowski:2012ih,McFadden:2013ria} it was found that the potential \eqref{srpot} has two local extremal points corresponding to those two CFTs, each of which reproduces the spectral tilt,
\begin{align}
(n_s-1)|_{\rm UV} = 2\lambda,\quad (n_s-1)|_{\rm IR} =- 2\lambda+\ldots~.
\end{align}
In this section, to discuss inflationary models with a red tilted spectrum (consistent with observations~\cite{Ade:2015xua}), we extend the holographic computation of the bispectrum to the IR region characterized by $\bar{\phi}k^{-\lambda}\gtrsim \lambda$, in contrast with \eqref{uvregion}.

\subsection{Inflationary bispectrum at the IR}

For the holographic computation of the bispectrum~\eqref{shape_function}, we need the two-point function
\begin{align}
\label{deformed_two-point}
\langle O_0(\boldsymbol k_1)O_0(\boldsymbol k_2)\rangle'& = \sum_{n=0}^\infty \frac{(-\bar\phi)^{n}}{n!} \langle O_0(\boldsymbol k_1)O_0(\boldsymbol k_2)O_0(\bsnl)^{n} \rangle'_{\rm CFT}\,,
\end{align}
and the three-point function
\begin{align}
\label{deformed_three-point}
\langle O_0(\boldsymbol k_1)O_0(\boldsymbol k_2)O_0(\boldsymbol k_3)\rangle'& = \sum_{n=0}^\infty \frac{(-\bar\phi)^{n}}{n!} \langle O_0(\boldsymbol k_1)O_0(\boldsymbol k_2)O_0(\boldsymbol k_3)O_0(\bsnl)^{n} \rangle'_{\rm CFT}\,,
\end{align}
in the deformed CFT. Just as we did in the previous section, we will compute these UV CFT correlation functions via the Witten diagrams in AdS$_4$ followed by the {replacement~\eqref{Mflip}}. In the IR region, higher-point CFT correlators become non-negligible, so it seems impossible to obtain the full correlators of the deformed CFT. Nevertheless, it turns out that the leading contributions in our parameter regime are calculable by performing a resummation over $\bar{\phi}k^{-\lambda}$. For concreteness (and as it turns out to be the most relevant for our purpose), we first summarize our main results of the resummation for the three-point function~\eqref{deformed_three-point}. The technical details are summarized in Appendices~\ref{wittendiagram} and \ref{explicit}.

\subsubsection{Outline of the computations}
The first goal is to evaluate all the CFT correlation functions in the expansions \eqref{deformed_two-point} and \eqref{deformed_three-point} at the leading order in $\lm$, at all orders in $g_3$ and up to the first order in the derivative couplings, following the strategy given in Sec.\,\ref{UVCFT}. Namely, starting from the AdS action \eqref{AdSaction}, we 
\begin{enumerate}[label=\arabic*)]
\item 
evaluate all Witten diagrams for the CFT correlation functions in \eqref{deformed_two-point} and \eqref{deformed_three-point} in our parameter regime,
\item substitute the results of Step 1) back into the expansions \eqref{deformed_two-point} and \eqref{deformed_three-point}, and carry out the summations keeping the condition $\bar{\phi}k^{-\lambda}\gtrsim \lambda$ in mind,
\item finally apply the {replacement \eqref{Mflip}} to obtain correlation functions of the perturbed~CFT.
\end{enumerate}
Step 1) seems intractable apparently, but actually the enumeration of the diagrams is simplified in our parameter region stated in Sec.\,\ref{infsetup}. First, we only need to take into account the cubic coupling $g_3$ out of the slow-roll potential because it dominates over higher-point couplings. Second, diagrams with two or more derivative couplings are subleading thanks to the conditions~\eqref{condition_alpha} on the derivative couplings.
These conditions on the model parameters enable us to complete the computation of relevant diagrams. The detail of Step 1) is given in Appendix \ref{wittendiagram}.
Then, in Step~2), we obtain integral representations of the full two-point and three-point functions after the summations. Remarkably, these integral expressions have very compact structures with what we call ``effective bulk-boundary propagators''. As explained in more detail shortly, they are bulk-boundary propagators ``dressed'' with zero-momentum propagators associated with the operator $O_0(\bsnl)$.
The result of Step 3) is given in Appendix \ref{adstods}.

\medskip
In the rest of this subsection we explain these results and their consequences more concretely: first evaluate the contributions to the three-point function without derivative couplings, which we will call the ``non-derivative part'' of the three-point function, and then evaluate the contributions at the first order in the derivative couplings, which we will call the ``$\alpha_m$-part''.

\subsubsection{Non-derivative part of the three-point function} \label{noderiv}
\begin{figure}[t]
  \centering
    \includegraphics[height=2in]{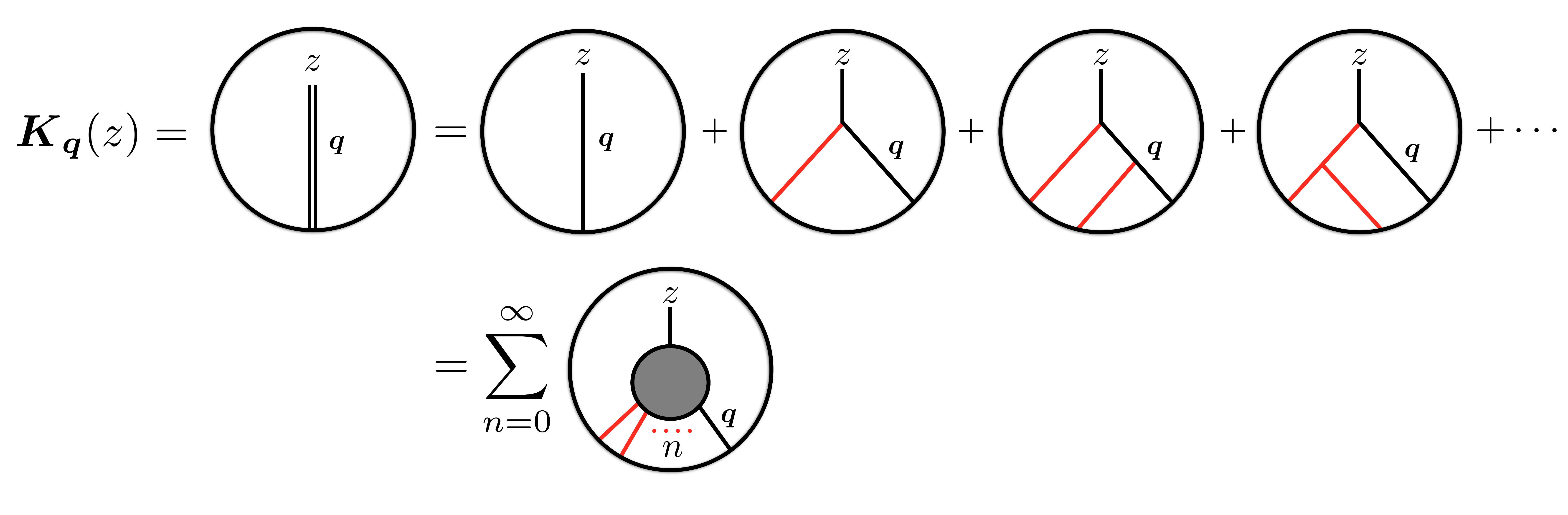}
    \caption{\label{fig:diagK} \small The effective bulk-boundary propagator is essentially taking all contributions of zero momentum legs (indicated in red). When the bulk point is pulled to the boundary according to \eqref{bbtobb}, it is exactly the two-point function of the perturbed CFT.}
\end{figure}

\begin{figure}[t]
  \centering
  \includegraphics[width=0.3\textwidth]{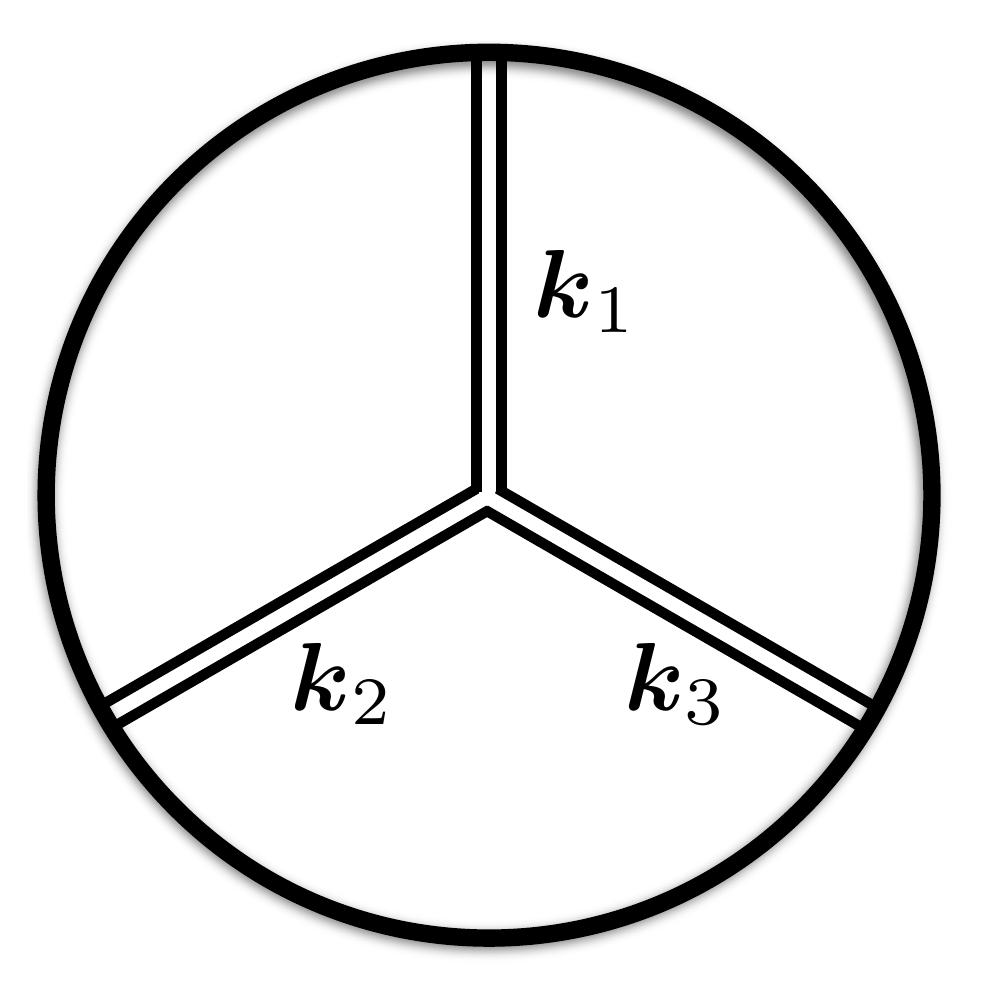} \quad\quad\quad\quad 
  \caption{\small The diagrammatic representation of the integral \eqref{effective_g_3}. The black double line denotes the effective bulk-boundary propagator $\bsK_{\boldsymbol k_i}(z)$.}\label{WittenDiagramOOOg3}
\end{figure}

Let us first focus on the non-derivative part of the three-point function \eqref{deformed_three-point}. 
As stated above, this part is contributed by Witten diagrams made up of three bulk-boundary propagators with $\boldsymbol k_i$ and an arbitrary number of zero-momentum bulk-boundary propagators, connected by the bulk-bulk propagators and the cubic vertex $g_3$.
We then introduce what we call the effective bulk-boundary propagator schematically as in Fig.\,\ref{fig:diagK}. It is essentially a bulk-boundary propagator dressed by zero-momentum propagators with the cubic vertex $g_3$ (see Appendix~\ref{wittendiagram} for a more precise definition). Notice that this dressing originates from the operator $O_0(\bsnl)$ in the perturbative expansion \eqref{deformed_three-point}.
All the relevant diagrams are now nicely reformulated into the form of Fig.\,\ref{WittenDiagramOOOg3}, where three effective bulk-boundary propagators are attached to the cubic coupling $g_3$. More explicitly,
\begin{align}
\label{effective_g_3}
\langle O_0(\boldsymbol k_1)O_0(\boldsymbol k_2)O_0(\boldsymbol k_3)\rangle' 
\ni 
-2g_3\int_0^\infty\frac{dz}{z^4}
{\bsK}_{\bsk_1}(z){\bsK}_{\bsk_2}(z){\bsK}_{\bsk_3}(z) \,,
\end{align}
where ${\bsK}_\bsq(z)$ denotes the effective bulk-boundary propagator of momentum $\bsq$.
This is exactly the integral expression of $\langle O_0(\boldsymbol k_1)O_0(\boldsymbol k_2)O_0(\boldsymbol k_3)\rangle_{\rm CFT}$ with the three bulk-boundary propagators replaced by the effective ones ${\bsK}_{\bsk_i}(z)$.

\medskip
Generally speaking, it is quite difficult to evaluate the integral~\eqref{effective_g_3}. However, it turns out to be somewhat tractable under the assumptions $\lambda\ll1$ and 
\begin{align}\label{approximation}
\bigg( \frac{k_1}{k} \bigg)^\lambda \sim \bigg( \frac{k_2}{k} \bigg)^\lambda\sim\bigg( \frac{k_3}{k} \bigg)^\lambda\sim 1\,.
\end{align}
where we introduced $k:=k_1+k_2+k_3$. From an inflationary point of view, the first assumption corresponds to the slow-roll condition as we mentioned earlier. On the other hand, the second one is essentially equivalent to the assumption  that the slow-roll parameters are approximately the same at the time of horizon crossing of each mode ${\bsk}_i$. In Appendix~\ref{slowroll}, we find that the integral~\eqref{effective_g_3} takes the form,
\begin{align}
f(\bar\phi k^{-\lambda}) (k_1^{3-3\lambda}+k_2^{3-3\lambda}+k_3^{3-3\lambda})\,,
\label{srshape}
\end{align}
where $f(\bar\phi k^{-\lambda})$ is some function irrelevant to the shape. Actually, because of some combinatorial complications, it seems not straightforward to determine the function $f(\bar\phi k^{-\lambda})$ directly.
Instead, in the following, we use the consistency relation \eqref{consistency_relation}, which we already proved on the CFT side for general setups, to find the explicit form of $f(\bar\phi k^{-\lambda})$.

\medskip
Let us start from the right hand side of the consistency relation~\eqref{consistency_relation}. The $\mathcal{O}(\alpha_m^0)$ part of the power spectrum and the spectral tilt were already computed in~\cite{vanderSchaar:2003sz,Bzowski:2012ih,McFadden:2013ria,Garriga:2014ema}, which gives
\begin{align}
P_\zeta = \frac{1}{4\pi^2\beta^2(\bar g(q))}+\mathcal{O}(\alpha_m)\,, \quad
n_s-1 = -2\beta'(\bar g(q)) \label{cftnsmo}+\mathcal{O}(\alpha_m)\,.
\end{align}
Here $\bar g(q)$ and $\beta (\bar g(q))$ are defined by
\begin{align}
\bar g(q) &:= 
\bar\phi q^{-\lambda} \bigg(1+\frac{g_3}{3}\frac{\bar\phi q^{-\lambda}}{\lambda}\bigg)^{-1}, \label{gbar} \\
\beta (\bar g(q)) &:= -\lambda \bar g(q)+\frac{g_3}{3} \bar g(q)^2 
= -\lambda\bar \phi q^{-\lambda}\bigg(1+\frac{g_3}{3}\frac{\bar\phi q^{-\lambda}}{\lambda}\bigg)^{-2}\,. \label{beta}
\end{align}
As discussed in~\cite{McFadden:2013ria}, $\bar{g}(q)$ and $\beta(g)$ may be identified with the coupling for the renormalized operator associated with $O_0$ and the corresponding beta function. Indeed, $\bar{g}(q)$ flows from UV to IR as
\begin{align}
0\leq\bar g(q)\leq \frac{3\lambda}{g_3}\,,
\end{align}
as expected from our inflationary setup. We may also see that $\beta$ is at most of the order $\lambda^2$ along the flow. On the other hand, the left hand side of Eq.~\eqref{consistency_relation} in the squeezed limit reads
\begin{align}
-2\lambda+\lambda\bar\phi\,\frac{{\rm Re}\langle   O_0 (\bsnl) O_0 ({\boldsymbol k}_2)  O_0 (-{\boldsymbol k}_2)  \rangle^\prime}{{\rm Re}\langle  O_0 ({\boldsymbol k}_2)  O_0 (-{\boldsymbol k}_2)  \rangle^\prime}
=-2\lambda - \frac{2(\lambda\bar{\phi}k^{-\lambda})^3f(\bar\phi k^{-\lambda})}{\bt(\bar g(k))^2}\,.
\end{align}
Here we used the approximation \eqref{approximation} and the two-point function,
\begin{align}
\lan O_0(\bsq)O_0(-\bsq) \ran'=-\frac{\bt(\bar g(q))^2}{(\lm\bar\phi)^2}q^3+\mathcal{O}(\alpha_m)\,, \label{two-point-fullnoalpha}
\end{align}
which follows from the result in~\cite{vanderSchaar:2003sz,Bzowski:2012ih,McFadden:2013ria,Garriga:2014ema}. The consistency relation \eqref{consistency_relation} then implies
\begin{align}
f(\bar\phi k^{-\lambda})
= -\Big(\lambda + \beta'(\bar g(k))\Big) \beta (\bar g(k))^2 (\lambda\bar\phi k^{-\lm})^{-3}
=-\frac{2g_3}{3\lambda}\bigg(1+\frac{g_3}{3}\frac{\bar\phi k^{-\lambda}}{\lambda}\bigg)^{-5}\,.
\end{align}

\subsubsection{$\alpha_m$-part of the three-point function} \label{alphamthree-point}
\begin{figure}[t]
  \centering
  \includegraphics[width=0.3\textwidth]{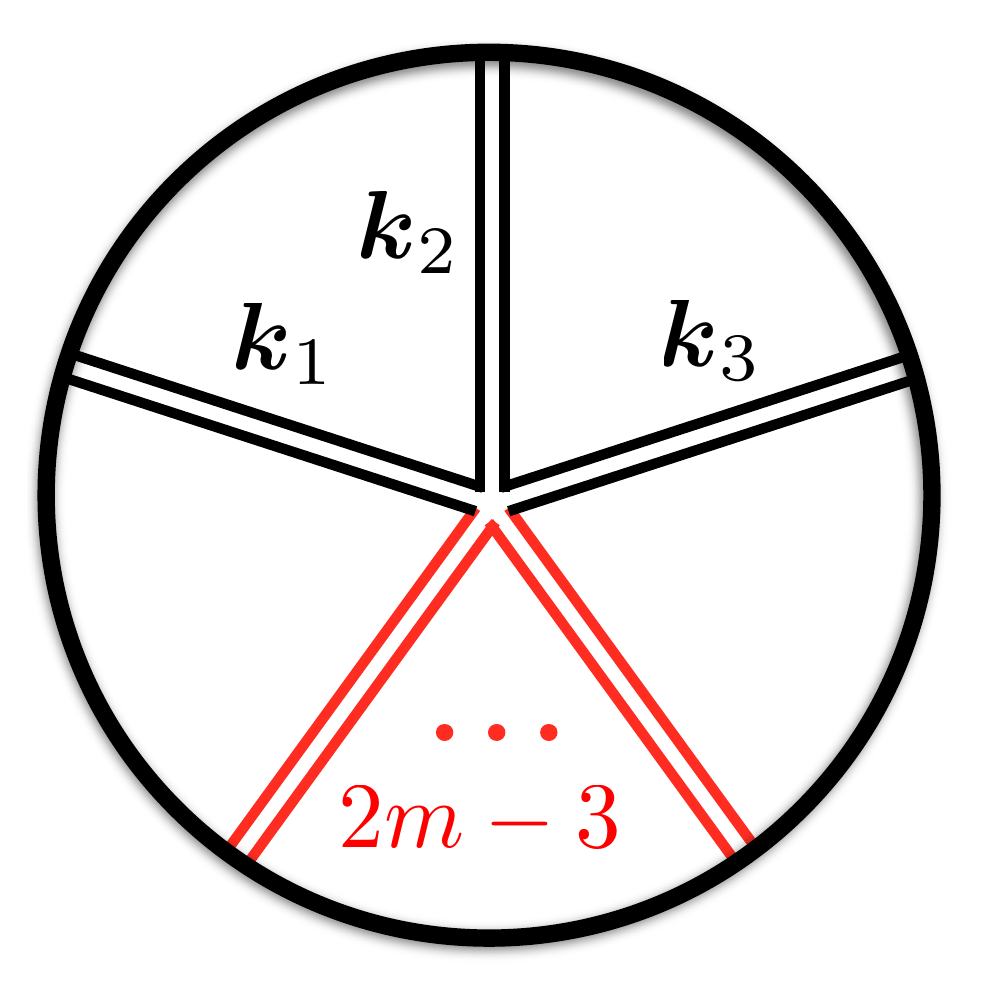}
  \caption{\small The diagrammatic representation of the integral \eqref{effective-an}. The red double line represents the effective zero-momentum bulk-boundary propagator $\bsE(z)$. 
  }\label{WittenDiagramOOOalpha}
\end{figure}

We next move on to the $\alpha_m$-part of the three-point function \eqref{deformed_three-point}. 
This part is contributed by all diagrams made up of three bulk-boundary propagators with with $\boldsymbol k_i$ and an arbitrary number of zero-momentum bulk-boundary propagators, connected by the bulk-bulk propagators, the cubic vertex $g_3$ and only one derivative coupling. Let us look at the diagrams with one $\alpha_m$. Similarly to the non-derivative case, all the relevant diagrams are nicely reformulated in terms of effective propagators as in Fig.\,\ref{WittenDiagramOOOalpha}.
Here, the new ingredient is the effective zero-momentum bulk-boundary propagator, i.e., the red double line in the figure. It is again dressed by zero-momentum propagators with the cubic vertex $g_3$ (see Appendix~\ref{wittendiagram} for more details). More explicitly, the integral representation of the $\alpha_m$-contribution to the three-point function~\eqref{deformed_three-point} is given by
\begin{align}
&\langle O_0(\boldsymbol k_1)O_0(\boldsymbol k_2)O_0(\boldsymbol k_3)\rangle' \nn\\
&\ni 
\sum_{m \geq 2} 
\frac{2m(2m-2)}{2^m}\alpha_m\int_0^\infty{dz} z^{2m-4}
\Big\{(2m-1)
{\bsK}_{\bsk_1}'(z){\bsK}_{\bsk_2}'(z){\bsK}_{\bsk_3}'(z){\bsE}'(z)^{2m-3}
\nn\\
& \quad\quad\quad\quad\quad\quad\quad\quad\quad
- \Big(\boldsymbol k_1.\boldsymbol k_2 {\bsK}_{\bsk_1}(z){\bsK}_{\bsk_2}(z){\bsK}_{\bsk_3}'(z){\bsE}'(z)^{2m-3} + (231)+(312)\Big) \Big\} \,,
\label{effective-an}
\end{align}
where ${\bsE}(z)$ denotes the zero-momentum effective bulk-boundary propagator. For the derivation, see  Appendix~\ref{corr}. This is exactly the integral expression of $\langle O_0(\boldsymbol k_1)O_0(\boldsymbol k_2)O_0(\boldsymbol k_3)O_0(\bsnl)^{2m-3}\rangle_{\rm CFT}$ for each $m$, with the three bulk-boundary propagators replaced by the effective ones ${\bsK}_{\bsk_i}(z)$ and $(2m-3)$ zero-momentum propagators replaced by the effective ones ${\bsE}(z)$.

\medskip
Surprisingly, it turns out that the assumption $\lambda\ll1$ drastically simplifies our computation. In Appendix~\ref{explicit} we show that the following approximations may be used in the integral \eqref{effective-an}:
\begin{align}
\label{simple_effective1}
\bsK_{\bsk_i}(z) 
&\simeq {\bigg( 1+\frac{g_3 \bar \phi k^{-\lambda}} {3\lambda} \bigg)^{-2} } \mathcal K_{\bsk_i}(z)\,, \\
\label{simple_effective2}
\bsK'_{\bsk_i}(z) 
&\simeq {\bigg( 1+\frac{g_3 \bar \phi k^{-\lambda}} {3\lambda} \bigg)^{-2} } \mathcal K'_{\bsk_i}(z)\,, \\
\label{simple_effective3}
\bsE'(z) 
&\simeq {\bigg( 1+\frac{g_3 \bar \phi k^{-\lambda}} {3\lambda} \bigg)^{-2} } \left[-\bar\phi \,\mathcal K'_{\bsnl}(z)\right]\,,
\end{align}
where $\mathcal K_{\bsq}(z)$ is the standard (undressed) bulk-boundary propagator,
\begin{align}
\cK_{\bsq}(z) = \frac{q^{(3/2)-\lm} z^{3/2}}{2^{(1/2)-\lm} \Ga((3/2)-\lm)} K_{(3/2)-\lm}(qz)\,. 
\end{align}
Interestingly, all the ingredients of the integral~\eqref{effective-an} have the same $z$-independent prefactor,
\begin{align}
\bigg( 1+\frac{g_3 \bar \phi k^{-\lambda}} {3\lambda} \bigg)^{-2}=\frac{\beta(\bar{g}(k))}{-\lambda \bar{\phi}k^{-\lambda}}\,,
\end{align}
which goes to unity in the UV limit $k\to\infty$.
As a result, the shape function takes a form quite similar to the UV results~\eqref{second_contribution}-\eqref{def_G}. By performing the integral \eqref{effective-an}, we have
\begin{align}
\langle O_0(\boldsymbol k_1)O_0(\boldsymbol k_2)O_0(\boldsymbol k_3) \rangle' \big|_{\al_m^1}
&=-4 k_1k_2k_3\frac{\beta(\bar g(k))^2}{(\lm\bar\phi)^3} \mathcal F\left(\frac{k_1}{k_3},\frac{k_2}{k_3}\right), 
\label{three-point-fullalpha}
\end{align}
with the shape function $\mathcal F(k_1/k_3,k_2/k_3)$ defined by
\begin{align}
\mathcal F\left(\frac{k_1}{k_3},\frac{k_2}{k_3}\right) 
&=
\left[-\frac{4}{3}X^2P_{XXX}\frac{3k_1k_2k_3}{2(k_1+k_2+k_3)^3} 
+2XP_{XX}G(k_1,k_2,k_3)\right]_{X=\bt(\bar g(k))^2/2}\,.
\end{align}
Here we used the approximations $\lm\ll1$ and (\ref{approximation}). Also notice that all the scale-dependence appears through $X=\bt(\bar g(k))^2/2$.

\medskip
Before stating the final result for the inflationary bispectrum, we would like to make some technical comments. After the computation of the integral~\eqref{effective-an}, one may wonder why the computation here was simpler than the one for the non-derivative part. In particular, one may wonder why we cannot use the approximations~\eqref{simple_effective1}-\eqref{simple_effective3} for the non-derivative contribution. Actually, the integrals \eqref{effective_g_3} and \eqref{effective-an} have a different behavior around $z=0$. First, the leading contribution of the latter integral is regular around $z=0$ essentially because $\alpha_m$'s are derivative interactions. We can therefore take the $\lambda\to0$ limit of the integrand before performing the integral. On the other hand, the former is singular for $\lambda\ll1$, so that we need to perform an analytic continuation of $\lambda$. Because of this analytic continuation, we cannot take the $\lambda\to0$ limit before performing the integral. Essentially, such a difference makes the computation in this subsection simpler compared with the non-derivative part. More details will be discussed in Appendix~\ref{explicit}.

\subsubsection{Final result}

We now compute the inflationary bispectrum using the obtained deformed QFT correlators. So far, we have computed the three-point function $\langle  O_0 ({\boldsymbol k}_1)  O_0 ({\boldsymbol k}_2)   O_0 ({\boldsymbol k}_3)  \rangle^\prime$ up to the first order in $\alpha_m$. Also, by evaluating the second integral of the right hand side of \eqref{full-twopos-dS}, we may show that the $\mathcal{O}(\alpha_m)$ correction to the two-point function $\lan O_0(\bsq)O_0(-\bsq) \ran'$ scales as $\sim\alpha_m\lambda^{4(m-1)}$, which is subdominant under the condition~\eqref{condition_alpha}. All in all, the leading contribution to the normalized bispectrum is given by 
\begin{align} \label{qftbi}
\frac{\langle \zeta({\boldsymbol k_1} )\zeta({\boldsymbol k_2}) \zeta({\boldsymbol k_3}) \rangle'}{\langle \zeta({\boldsymbol k_1}) \zeta({-\boldsymbol k_1}) \rangle' \langle \zeta({\boldsymbol k_2}) \zeta({-\boldsymbol k_2}) \rangle' }
&= \bt'(\bar g(k)) \bigg( 1+\!\!\left(\frac{k_1}{k_3}\right)^3+\!\!\left(\frac{k_2}{k_3}\right)^3 \bigg) 
+\frac{4k_1k_2k_3}{k_2^3}  \mathcal F\left(\frac{k_1}{k_3},\frac{k_2}{k_3}\right).
\end{align}

\subsection{Comparison with inflationary results} \label{RG}
In this subsection, we compare the result \eqref{qftbi} from the dual QFT with the result on the inflation side. For this we first relate the coupling $\bar g$ \eqref{gbar} with the inflaton.
The coupling $\bar g$ satisfies 
\begin{align}\label{CFTsideg}
\frac{d \bar g(k)}{d\ln k} = \bt(\bar g(k)) = -\lambda \bar g(k) + \frac{g_3}{3} \bar g^2(k).
\end{align}
On the inflation side, the time derivative of the inflaton trajectory $\phi_c(t)$ can be written in terms of the inflaton trajectory $\phi_c$ itself as\footnote{This is derived by writing $\dot\phi_c$ in terms of $\phi_c$. See \cite{McFadden:2013ria} for the detail for a general slow-roll potential.}
\begin{align}\label{inflationsidephi}
\dot \phi_c = -\lambda \phi_c+\frac{g_3}{3} \phi_c^2 + \cO(\al_m).
\end{align}
This equation holds at any time, especially at the horizon crossing $t_*(k)$ satisfying $k = aH(t_*(k))$. Comparing (\ref{CFTsideg}) and (\ref{inflationsidephi}), we can identify the running coupling $\bar g(k)$ at the leading order in $\lambda$ with the inflaton at the horizon exit $\phi_c(t_*(k))$ up to $\al_m$-corrections,
\begin{align} \label{identify}
\phi_c(t_*(k)) = \bar g(k)+\cO(\al_m).
\end{align}

Let us apply this identification to the bispectrum directly computed on the inflation side \cite{Chen:2006nt}. In our parameter regime, the leading contribution is
\begin{align}
&\frac{\langle \zeta({\boldsymbol k_1} )\zeta({\boldsymbol k_2}) \zeta({\boldsymbol k_3}) \rangle'}{\langle \zeta({\boldsymbol k_1}) \zeta({-\boldsymbol k_1}) \rangle' \langle \zeta({\boldsymbol k_2}) \zeta({-\boldsymbol k_2}) \rangle' } \nn\\[+5pt]
&\quad =
\frac{\eta}{2}\bigg( 1+\!\!\left(\frac{k_1}{k_3}\right)^3+\!\!\left(\frac{k_2}{k_3}\right)^3 \bigg)  \nn\\[+5pt]
& \quad\quad
+\frac{4k_1 k_2 k_3}{k_3^3}\bigg[\left(\frac{1}{c_s^2} -1 -2\frac{\Lambda}{\Sigma} \right) \frac{3 k_1 k_2 k_3}{2(k_1+k_2+k_3)} 
+\left(\frac{1}{c_s^2} -1\right) G(k_1,k_2,k_3)
\bigg],
\label{bispectrum-inflation}
\end{align}
where the parameters $\Lm, \Sig$ are defined as \cite{Chen:2006nt}
\begin{align}
\Lm:=XP_{XX}+\frac{2}{3}X^2P_{XXX}, \quad
\Sig:=P_X+2XP_{XX}.
\end{align}
Here these parameters and the sound speed are defined with $X=\frac{1}{2}\dot\phi_c^2$.
First the slow-roll parameter $\eta$ becomes
$\eta=2\bt'$. 
Next, the remaining parameters become, at the leading order in the derivative couplings,
\begin{align}
X &= \frac{1}{2}\dot\phi_c^2\simeq\frac{1}{2}\beta(\bar g(k))^2, \nn\\[+5pt]
\frac{1}{c_s^2}-1 &= \frac{2XP_{XX}}{P_X} \simeq 2XP_{XX}, \nn\\[+5pt]
\frac{1}{c_s^2}-1-\frac{2\Lm}{\Sigma} &= \frac{2XP_{XX}}{P_X}-2\frac{XP_{XX}+\frac{2}{3}X^2P_{XXX}}{P_X+2XP_{XX}} \simeq -\frac{4}{3}X^2P_{XXX},
\end{align}
where $\simeq$ stands for the equality up to the leading order in $\al_m$.
Applying these results to \eqref{bispectrum-inflation}, we find it exactly coincides with \eqref{qftbi}, as desired.

\section{Summary and Outlook}
\setcounter{equation}{0}

In this paper, we applied conformal perturbation  techniques to holographically compute the bispectrum of a generalized single-field inflation model containing derivative couplings. Correlation functions of the reference UV CFT were computed via the dS/CFT correspondence. Using them, we first computed correlation functions of a relevantly deformed CFT perturbatively around the UV fixed point, and obtained a blue tilted power spectrum as well as a bispectrum of the correct shape. Then, by taking into account all-order conformal perturbations to reach the IR region of the RG flow, we reproduced the bispectrum up to first order in the derivative couplings at scales where the power spectrum is red tilted.

\medskip
We would like to end this paper with several promising future directions. First, our analysis in the present paper was perturbative in the derivative couplings. While our result is exact to all orders in the cubic coupling $g_3$, it is valid only around the parameter region with $c_s^2\sim 1$. It would be interesting to generalize our findings to the small sound speed case where $c_s^2 \ll 1$. Another interesting direction will be to extend our results to  quasi single-field inflation~\cite{Chen:2009zp,Baumann:2011nk,Noumi:2012vr} (which contains massive fields in addition to the inflaton), especially in connection to the ``cosmological collider physics"  program \cite{Arkani-Hamed:2015bza}. The CFT viewpoint will be useful in understanding various soft limit properties of the inflationary correlators. Furthermore, 
recent developments in higher dimensional CFT, such as the conformal bootstrap approach,\footnote{ \Red{See, e.g.,~\cite{Gliozzi:2013ysa,Gliozzi:2014jsa} for application of the conformal bootstrap approach to non-unitary CFTs, which will be relevant in the context of dS/CFT.}}
may give us additional handles to derive inflationary non-Gaussianities.
We expect such technical advances, together with conformal perturbation theory and holography that we exploited in this paper, would shed new light on the structure of primordial spectra. We hope to report our progress in these directions elsewhere in the near future.

\section*{Acknowledgement}
\noindent
We thank Junyu Liu for initial collaboration during his internship at the Hong Kong University of Science and Technology. We thank Ignatios Antoniadis, Auttakit Chatrabhuti, Xingang Chen, Oleg Evnin, Hongliang Jiang, Juan Maldacena, Razieh Emami Meibody, Yutaka Ookouchi, Yuki Sato, Shigeki Sugimoto, Henry Tye and Yi Wang for useful discussions.
This work is supported in part by the Research Grants Council (RGC) of Hong Kong through  grants HKUST4/CRF/13G and 16304414.
HI is supported by the ``CUniverse'' research promotion project by Chulalongkorn University (grant reference CUAASC).
GS is supported in part by the DOE grant DE-FG-02-95ER40896 and the Kellett Award of the University of Wisconsin.
SZ is supported by the Hong Kong PhD Fellowship Scheme (HKPFS) issued by the RGC of Hong Kong. 
We also thank the Yukawa Institute for Theoretical Physics at Kyoto University for hospitality during the workshop YITP-W-16-05 ``Strings and Fields 2016'', where HI presented the results in this paper. He benefited from the fruitful discussion there.

\appendix

\section{Correlation functions from AdS}\label{wittendiagram}

This and the next appendices will give a detailed account of computing CFT correlation functions from an AdS action. In this appendix, we give a classical solution to the equation of motion of the AdS action perturbatively, compute their derivatives and use them to write down the correlation functions of our interest. See the appendix B of \cite{McFadden:2013ria} for a nice review for the canonical kinetic term case. This section is its extension to the case with derivative couplings.

\setcounter{equation}{0}

\subsection{Review of holographic computations in AdS}
The bulk AdS action of our interest is
\begin{align}
S_{\rm AdS}=\int d^4x\sqrt{G} \left[
\frac{1}{2}(\pd\Phi)^2+\frac{m^2}{2}\Phi^2-\frac{g_3}{3}\Phi^3 
+\sum_{m \geq 2}\frac{a_m}{2^m}(\pd\Phi)^{2m}
\right], 
\label{AdSmodel}
\end{align}
where $(\pd\Phi)^2 := G^{MN}\pd_M\Phi\pd_N\Phi$ and $G_{MN}$ is the Euclidean four-dimensional AdS metric $ds^2=R_{\rm AdS}^2z^{-2}(dz^2+d\bsx^2)$.
The equation of motion of $\Phi$ is
\begin{align}
- G^{-1/2}\pd_M(G^{1/2}\pd^M\Phi)
+ m^2\Phi-g_3\Phi^2 
- \sum_{m \geq 2}\frac{a_m}{2^m}{G^{-1/2}}\pd_M(G^{1/2}\pd^M\Phi (\pd\Phi)^{2(m-1)}) 
= 0.
\end{align}
Let us introduce a parameter $\nu$ and formally expand $\Phi$ in $\nu$ as $\Phi=\sum_{n=1}^\infty\nu^{n}\Phi_{n}$. 
Then the equations of motion for $\Phi_{n}$ read
\begin{align}
0 &= \cD\Phi_{1}, \label{eomzq1} \\
0 &= \cD\Phi_{2}-g_3\Phi_{1}^2, \label{eomzq2} \\
0 &= \cD\Phi_{n}-g_3\sum_{\substack{a+b=n \\ a,b \geq 1}}\Phi_a\Phi_b
-\sum_{m \geq 2}\frac{2ma_m}{2^m}\!\!\!\!
\sum_{\substack{a+\sum_i(b_i+c_i)=n \\ a,b_i,c_i \geq 1}} \!\!\!\!\!\!
G^{-1/2}\pd_M(G^{1/2}\pd^M\Phi_a(\pd_N\Phi_{b_i}\pd^N\Phi_{c_i})^{m-1}), \label{eomzqn}
\end{align}
where the third equation is for $n \geq 3$, and $\cD$ is the differential operator
\begin{align}
\cD\Phi := - G^{-1/2}\pd_M(G^{1/2}\pd^M\Phi) + m^2\Phi.
\end{align}
We solve the first equation of motion \eqref{eomzq1} by
\begin{align}\label{Phi1}
\Phi_1(X) = \int_\bsx\cK_X(\bsx)\vphi_{[0]}(\bsx),
\end{align}
where $\vphi_{[0]}(\bsx)$ is the boundary field, $X=(z,\bsy)$ denotes a bulk point, and 
we introduced the bulk-boundary propagator $\cK_X(\bsx)$, which satisfies $\cD_X\cK_X(\bsx)=0$. We will use its three-dimensional momentum expression
\begin{align}
\cK_{\bsq}(z) := \frac{q^{(3/2)-\lm} z^{3/2}}{2^{(1/2)-\lm} \Ga((3/2)-\lm)} K_{(3/2)-\lm}(qz), \quad
R_{\rm AdS}^2m^2 = \lm(\lm-3),
\end{align}
which is defined by
\[
\cK_{(z,\bsy)}(\bsx)=\int \frac{d^3q}{(2\pi)^3} e^{i\bsq.(\bsx-\bsy)}\cK_\bsq(z).
\]
Here the normalization factor in $\cK_{\bsq}$ is fixed such that $\cK_{\bsq}(z) \to z^\lm$ as $z \to 0$.
To solve the remaining equations of motion \eqref{eomzq2} and \eqref{eomzqn}, we introduce the Green's function $\cG(z,\bsx;z',\bsx')$, called the bulk-bulk propagator, as the solution to the free equation of motion with a delta-function source
\begin{align}
R_{\rm AdS}^2\cD_z\cG(z,\bsx;z',\bsx')=z^{4}\de(z-z')\de^3(\bsx-\bsx').
\end{align}
The overall $R_{\rm AdS}^2$ on the left hand side is introduced so that the $R_{\rm AdS}^2$-dependence only comes from the mass in the form $R_{\rm AdS}^2m^2$.
Its three-dimensional momentum space $\cG_{\bsq}(z,z')$ reads 
\begin{align}
\cG_{\bsq}(z,z') =
\begin{cases}
(zz')^{3/2} K_{\frac{3}{2}-\lm}(qz) I_{\frac{3}{2}-\lm}(qz') &\quad z>z', \\
(zz')^{3/2} I_{\frac{3}{2}-\lm}(qz) K_{\frac{3}{2}-\lm}(qz') &\quad z'>z,
\end{cases}
\end{align}
which is defined by
\[
\cG(z,\bsx;z',\bsx')=\int \frac{d^3q}{(2\pi)^3} e^{i\bsq.(\bsx-\bsx')}\cG_\bsq(z,z').
\]
The solutions to the equations of motion \eqref{eomzq2} and \eqref{eomzqn} then read
\begin{align}
\Phi_{2,X} &= g_3R_{\rm AdS}^{-2}\int_{X'}\cG_{XX'}\Phi_{1,\bsX'}\Phi_{1,\bsX'}, \label{Phi2} \\
\Phi_{n,X} &= 
g_3R_{\rm AdS}^{-2}\int_{X'}\cG_{XX'}
\sum_{\substack{a+b=n \\ a,b \geq 1}}\Phi_{a,X'}\Phi_{b,X'} 
\nn\\
&{} 
+\sum_{m \geq 2}\frac{2m}{2^m}a_mR_{\rm AdS}^{2(m-5)}\int_{X'} \cG_{XX'} \!\!\!\!\!\!\!\!
\sum_{\substack{a+\sum_i(b_i+c_i)=n \\ a,b,c \geq 1}} \!\!\!\!\!\!\!\! z'^{4} 
\pd'_M\left[ z'^{2(m-2)}\pd'_M\Phi_{a,X'}(\pd'_N\Phi_{b_i,X'}\pd'_N\Phi_{c_i,X'} )^{m-1}
\right]\,, \label{Phin}
\end{align}
where each of the indices $M,N$ is summed over $z,\bsx$.\footnote{Note that the integral measure also involves $R_{\rm AdS}^{4}$ from $\sqrt{G}$, which, however, is irrelevant to the flip \eqref{Mflip}.}
Just for notational simplicity, we introduced the abbreviations: for 
$ \Phi_{a,X}=\Phi_a(z,\bsx), \ \cG_{XX'}=\cG(z,\bsx;z',\bsx'), \ G_{z}^{MN}=G^{MN}(z), \pd_M'=\frac{\pd}{\pd z'^M} $ and  $ \int_{X'} := \int\frac{dz'}{z'^4}\int {d^3\bsx'} $.

\begin{figure}[htbp]
  \centering
    \includegraphics[height=1.2in]{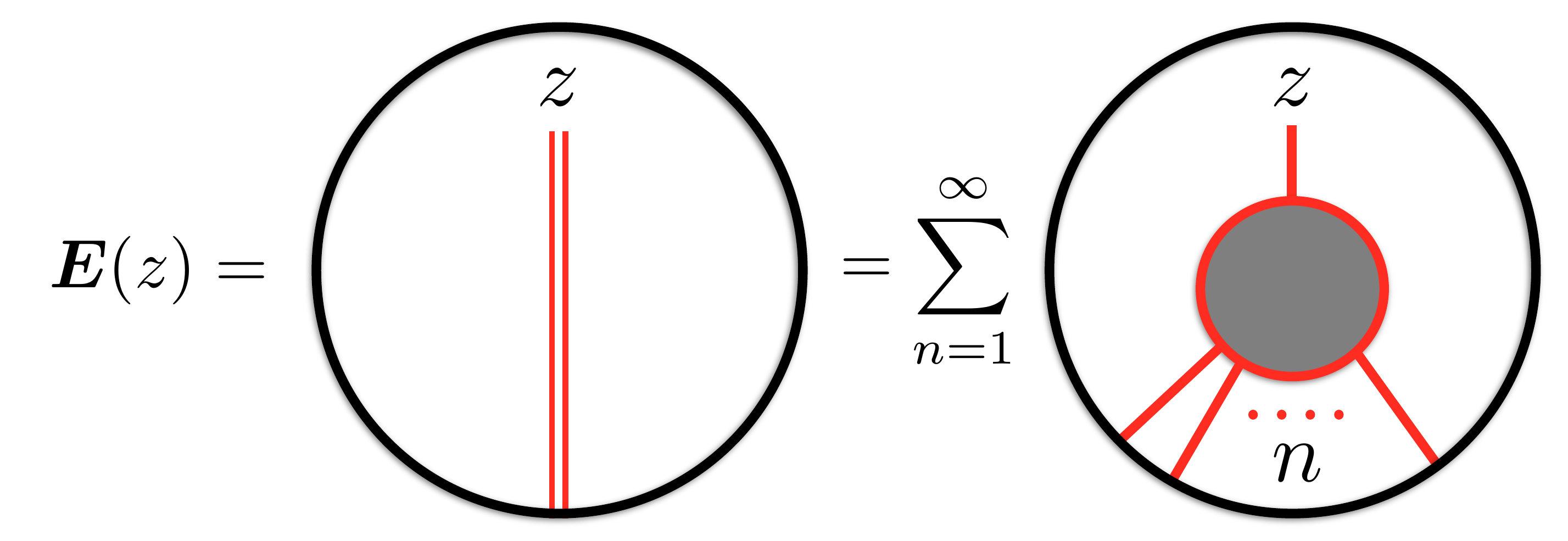} \quad\quad
    \includegraphics[height=1.2in]{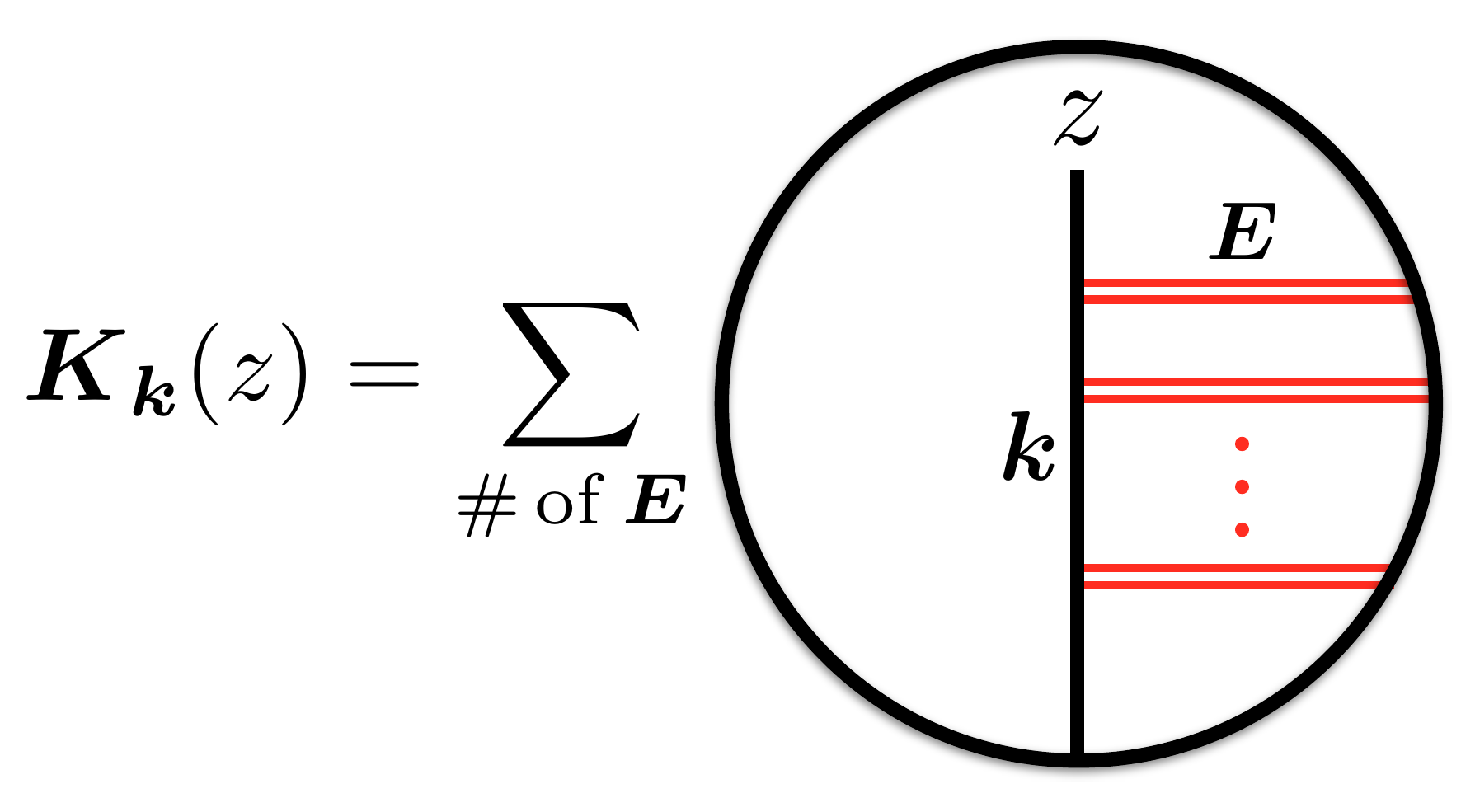}
  \caption{\label{fig:diagE} \small The diagram on the left is the Witten diagram of the effective zero momentum bulk-boundary propagator $\bsE(z)$. The one on the right is the effective bulk-boundary propagator $\bsK_{\bsk}(z)$ in terms of $\bsE$. See also Fig.\,\ref{fig:diagK}.}
\end{figure}

\paragraph{AdS/CFT dictionary} To describe the AdS/CFT correspondence precisely, we use the asymptotic behavior of $\Phi$ as $z \to 0$, which is given as the sum of two power series
\begin{align}
\Phi(z) = 
z^{\lm}\left( \vphi_{[0]}+z^2\vphi_{[2]}+\cdots \right)
+z^{3-\lm}\left( \vphi_{[3-2\lm]}+z^2\vphi_{[5-2\lm]}\cdots \right).
\end{align}
The holographic dictionary is to identify $\vphi_{[3-2\lm]}$ with the 1-point function in the presence of the boundary field $\vphi_{[0]}$ up to contact terms \cite{deHaro:2000vlm},
\begin{align} \label{Ophi}
\lan O_0 \ran_{\rm s} = -R_{\rm AdS}^2(3-2\lm)\vphi_{[3-2\lm]}.
\end{align}
One can extract $\vphi_{[3-2\lambda]}$ from \eqref{Phin} by expanding $\cG_{XX'}$ at $z \to 0$, $z<z'$,
\begin{equation}\label{bbtobb}
  \cG_{\bsq}(z,z') = \frac{1}{3-2\lambda}\cK_{\bsq}(z') z^{3-\lm}+ {\cal O}(z^{5-\lm}),
\end{equation}
such that the contribution of $\Phi_{n,X}$ to $\langle O_0\rangle_s$ are obtained by replacing $\cG_{\bsq}(z,z')$ by $\cK_{\bsq}(z')$ in \eqref{Phin}. 
Correlation functions are obtained by taking derivatives with respect to $-\vphi_{[0]}$
\begin{align} \label{Ocorrphi}
\lan O_0(\bsk_1) ... O_0(\bsk_n) \ran = (-)^{n-1}\frac{\de^{n-1}\lan O_0(\bsk_1) \ran_{\rm s}}{\de\vphi_{[0]}(-\bsk_2) ... \de\vphi_{[0]}(-\bsk_n)}\Big|_{\vphi_{[0]}=0}.
\end{align}
Note that the correlation functions thus obtained are those of CFT$_3$ dual to AdS$_4$.
We can see from \eqref{Phi1}, \eqref{Phi2} and \eqref{Phin} that $\Phi_n$ contains $n$ source fields $\vphi_{[0]}$. They can be drawn as Witten diagrams with the bulk-bulk propagator $\cG$ and the bulk-boundary propagator $\cK$ and all the combinatorial factors are contained in $\Phi_n$.
Correlation functions in our perturbed CFT are expressed as an infinite sum of the CFT correlators, for instance,
\begin{align}
\lan O_0(\bsk_1)O_0(\bsk_2) \ran'=\sum_{n=0}^\infty\frac{(-\bar\phi)^n}{n!}
\lan O_0(\bsk_1)O_0(\bsk_2)O_0(\boldsymbol 0)^n\ran'_{\rm CFT}. 
\end{align}
Since all UV CFT correlation functions contribute at the same order at the IR, we need $n$-point functions for general $n$ in which two or three operators have nonzero momenta and others have zero momentum. To evaluate them, we first take the $n$-th, $(n-1)$-th, and $(n-2)$-th derivatives of $\Phi_n$ with respect to the boundary fields with zero momentum. Moreover, we will evaluate the correlation functions to all orders in $g_3$ and to first order in $a_m$. We will finally find that the two-point and three-point functions can be compactly written in terms of the effective bulk-boundary propagators, which can be depicted as Witten diagrams in Fig.\,\ref{WittenDiagramOOOg3} plus Fig.\,\ref{WittenDiagramOOOalpha}. Diagrammatic representation of the effective bulk-boundary propagators will be given in Fig.\,\ref{fig:diagK} and \ref{fig:diagE}.

For later convenience, we first decompose $\Phi_n$ as
\begin{align}
\Phi_n=\Phi^{(g_3)}_n+\sum_{m \geq 2}\Phi^{(a_m)}_n.
\end{align}
Here $\Phi^{(g_3)}_n$ has no derivative couplings $a_m$ while $\Phi^{(a_m)}_n$ has only one $a_m$ for each $m$.

\paragraph{Recovering the AdS radius}
From now on, we will set $R_{\rm AdS}=1$ for simplicity. We can recover the $R_{\rm AdS}^2$ dependence by applying the following replacement
\begin{align} \label{rads}
g_3 \longrightarrow R_{\rm AdS}^{-2}g_3, \quad a_m \longrightarrow R_{\rm AdS}^{2(m-5)}a_m, \quad
\lan OO... \ran_{\rm AdS} \longrightarrow R_{\rm AdS}^2 \lan OO... \ran_{\rm AdS}.
\end{align}

\subsection{Properties of $\Phi^{(g_3)}_{n}$}
Keeping the correlation functions of our interest in mind, we introduce
\begin{align}
C^{(g_3)}_{n,\bsk_1,\bsk_2}(z)
&= 
\frac{2}{(n-2)!(2g_3)^{n-1}}
\frac{\de^{n}\Phi^{(g_3)}_{n}(z,\bsk_3)}
{\de\vphi_{[0]}(-\bsk_1)\de\vphi_{[0]}(-\bsk_2)\de\vphi_{[0]}(\bsnl)^{n-2}}, 
\label{Cg} \\
K^{(g_3)}_{n,\bsk_1}(z)
&= 
\frac{1}{(n-1)!(2g_3)^{n-1}} 
\frac{\de^{n}\Phi^{(g_3)}_{n}(z,\bsk_2)}{\de\vphi_{[0]}(-\bsk_1)\de\vphi_{[0]}(\bsnl)^{n-1}}, 
\label{Kg} \\
E^{(g_3)}_{n}(z)
&= 
\frac{1}{n!(2g_3)^{n-1}} \frac{\de^{n}\Phi^{(g_3)}_{n}(z,\bsnl)}{\de\vphi_{[0]}(\bsnl)^{n}}, 
\label{Eg} 
\end{align}
where we have omitted momentum-conserving delta function $(2\pi)^3\delta(\sum \bsk_i)$.\footnote{More concretly, for \eqref{Cg}, it holds that $\bsk_1+\bsk_2+\bsk_3=\bsnl$, and for \eqref{Cg}, it holds that $\bsk_1+\bsk_2=\bsnl$.} Here we make a comment on the meaning of the normalizations above. We will shortly find that the recursion relations for $C^{(g_3)},K^{(g_3)},E^{(g_3)}$ involve no coupling constant $g_3$. This implies that the $C^{(g_3)},K^{(g_3)},E^{(g_3)}$ are all independent of $g_3$, so that the normalizations above tell us the exact $g_3$-dependence of the above three derivatives of $\Phi^{(g_3)}_n$.

\begin{figure}[htbp]
\centering
\includegraphics[height=1.2in]{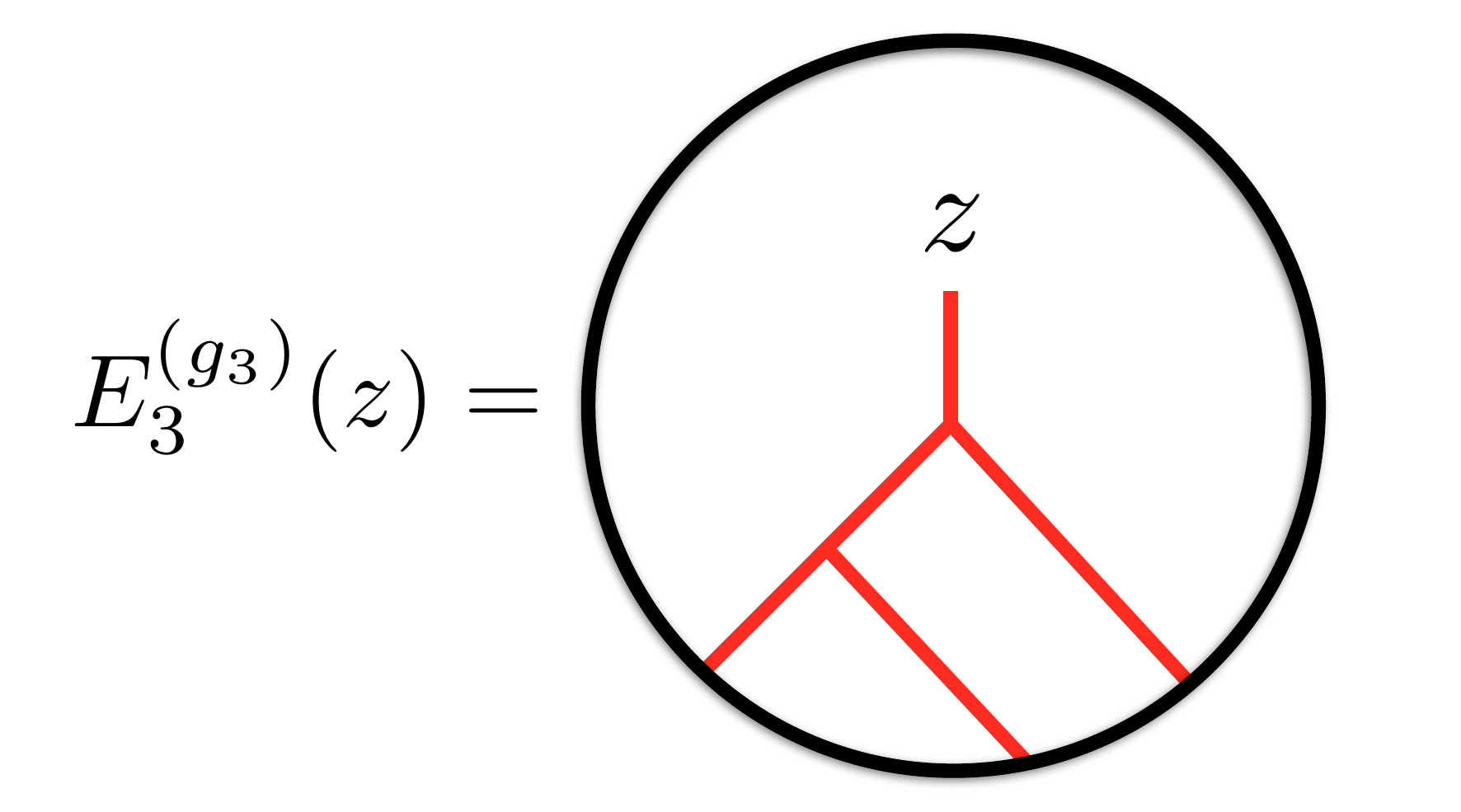}
\includegraphics[height=1.2in]{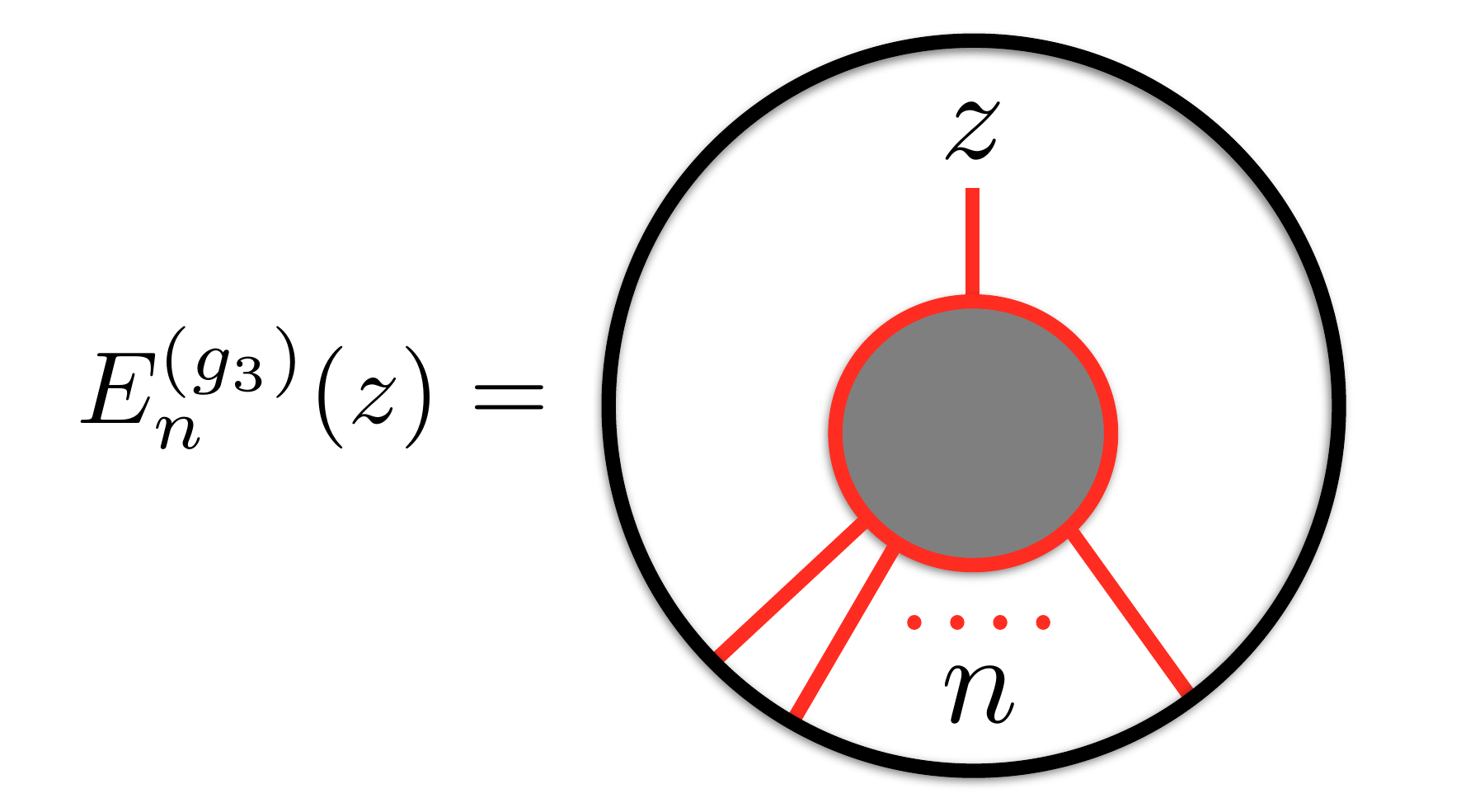}
\caption{\label{Ediagram} \small Diagrams for $E^{(g_3)}_{n}$. The index $n$ is the number of the zero-momentum bulk-boundary propagators.}
\end{figure}

Taking derivatives of \eqref{Phi1}, \eqref{Phi2} and \eqref{Phin}, we find the recursion relations for $E^{(g_3)}_n$
\begin{align}
E^{(g_3)}_{1}(z)&=\cK_{\bsnl}(z), \label{Eg1} \\
E^{(g_3)}_{n}(z)&=
\frac{1}{2}\int_{z'}\sum_{\substack{a+b=n \\ a,b \geq 1}} 
\cG_{\bsnl}(z,z')E^{(g_3)}_{a}(z')E^{(g_3)}_{b}(z'). \quad (n \geq 2) \label{Egn}
\end{align}
This can be depicted as Fig.\,\ref{Ediagram}

The recursion relations for $K^{(g_3)}_n$ read
\begin{align}
K^{(g_3)}_{1,\bsk}(z)&= \cK_{\bsk}(z), \label{Kg1} \\
K^{(g_3)}_{n,\bsk}(z)&=
\int_{z'}\sum_{\substack{a+b=n \\ a,b \geq 1}} 
\cG_{\bsk}(z,z')E^{(g_3)}_{a}(z')K^{(g_3)}_{b,\bsk}(z'). \quad (n \geq 2) \label{Kgn}
\end{align}
In terms of $E^{(g_3)}$, it reads,
\begin{align} \label{rec-Kn}
K^{(g_3)}_{n,\bsk}(z)
=\!\!\!\!\!\!\!\!
\sum_{\substack{p\geq0 \\n_1+...+n_p=n-1 \\ 1 \leq n_1,...,n_p \leq n-1}} \!\!\!\!\!\!\!\!
\int_{z_1,...,z_p} \!\!\!\!\!\!\!\!
\cG_{\bsk}(z,z_1)E^{(g_3)}_{n_1}(z_1)\cG_{\bsk}(z_1,z_2)E^{(g_3)}_{n_2}(z_2)\cdots\cG_{\bsk}(z_{p-1},z_p)E^{(g_3)}_{n_p}(z_p)\cK_{\bsk}(z_p).   
\end{align}
The summation symbol $\sum_{p \geq 0}\sum_{{n_1+...+n_p=n-1, 1 \leq n_1,...,n_p \leq n-1}}$ means that the sum over all possible decompositions of $n-1$ into a sum of integers no less than 1.
In the $n=1$ case, this decomposition does not exist, namely $p=0$, consistent with \eqref{Kg1}. Diagrammatically, it simply means attaching $(n-1)$ zero momentum legs to the bulk-boundary propagator using the $g_3$ vertex. 

The recursion relations for $C^{(g_3)}_{n}$ read
\begin{align}
C^{(g_3)}_{2,\bsk_1,\bsk_2}(z)&= \int_{z'} \cG_{\bsk_1+\bsk_2}(z,z') K^{(g_3)}_{1,\bsk_1}(z')K^{(g_3)}_{1,\bsk_2}(z'), \label{Cg2}\\
C^{(g_3)}_{n,\bsk_1,\bsk_2}(z)&=
\int_{z'}\sum_{\substack{a+b=n \\ a,b \geq 1}} \cG_{\bsk_1+\bsk_2}(z,z')\left[E^{(g_3)}_{a}(z')C^{(g_3)}_{b,\bsk_1,\bsk_2}(z')+
K^{(g_3)}_{a,\bsk_1}(z')K^{(g_3)}_{b,\bsk_2}(z') \right]. \quad (n \geq 3)
\end{align}
It can be expressed as
\begin{align}
C^{(g_3)}_{n,\bsk_1,\bsk_2}(z)
=\int_{X'}\sum_{\substack{a+b+c=n \\ a \geq 0, ~ b,c \geq 1}}
G_{a,\bsk_1+\bsk_2}(z,z')K^{(g_3)}_{b,\bsk_1}(z')K^{(g_3)}_{c,\bsk_2}(z'),
\label{Cg3n}
\end{align}
where we defined for $n \geq 0$
\begin{align}
&{}
G_{n,\bsk}(z,z'):= \nn\\
&{} \quad
\sum_{ \substack{p \geq 0\\ n_1+...+n_p=n \\ 1 \leq n_1,...,n_p \leq n}} \!\!\!\!\!\!
\int_{z_1,...,z_p}
\cG_{\bsk}(z,z_1)E^{(g_3)}_{n_1}(z_1)\cG_{\bsk}(z_1,z_2)E^{(g_3)}_{n_2}(z_2)\cdots
\cG_{\bsk}(z_{p-1},z_p)E^{(g_3)}_{n_p}(z_p) \cG_{\bsk}(z_p,z'),
\end{align}
where $G_{0,\bsk}$ is equal to $\cG_{\bsk}$. The following figure shows this definition diagrammatically:
\begin{figure}[htbp]
  \centering
    \includegraphics[height=1.2in]{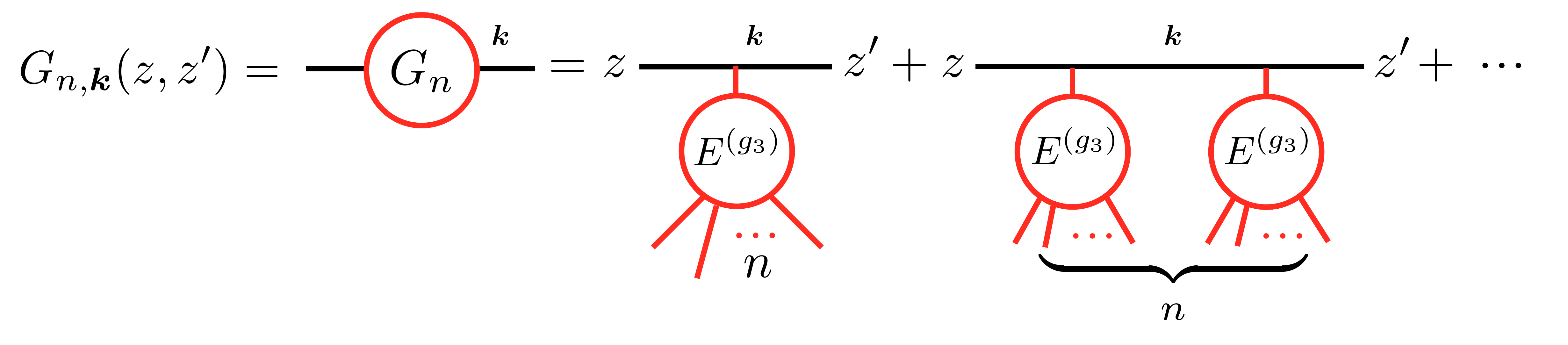}
\end{figure}

\subsection{Properties of $\Phi_n^{(a_m)}$}
In a similar manner, from $\Phi^{(a_m)}_n$ one can obtain $C^{(a_m)}_n$ with $(n-2)$ zero momentum legs and one $a_m$ vertex, $K^{(a_m)}_n$ with $(n-1)$ zero momentum legs and one $a_m$ vertex, and $E^{(a_m)}_n$ with $n$ zero momentum legs and one $a_m$ vertex. Here we explain the case $m=2$ in detail. The extension to general $m$ is straightforward, though the computations will be a bit more complicated. For instance, 
the three-point function is derived from $C^{(a_2)}_{n,\bsk_1,\bsk_2}$,
\begin{align}
C^{(a_2)}_{n,\bsk_1,\bsk_2}(z)
&:= 
\frac{1}{(n-2)!a_2(2g_3)^{n-3}} 
\frac{\de^{n}\Phi^{(a_2)}_{n}(z,\bsk_3)}
{\de\vphi_{[0]}(-\bsk_1)\de\vphi_{[0]}(-\bsk_2)\de\vphi_{[0]}(\bsnl)^{n-2}} \nn\\[+5pt]
& \!\!\!\!\!\!\!\!\!\!\!\!\! =
-\int_{z'}  \sum_{\substack{a+b+c+d=n \\ a \geq 0, ~ b,c,d \geq 1}} \!\!\!\!
\pd'_MG_{a,\bsk_1+\bsk_2}(z,z')
\Big[ 
\pd'^ME^{(g_3)}_{b}(z')\pd'_NK^{(g_3)}_{c,\bsk_1}(z')\pd'^NK^{(g_3)}_{d,\bsk_2}(z') \nn\\[-15pt]
&{} \quad\quad\quad\quad\quad\quad\quad\quad\quad\quad\quad\quad\quad\quad
+2\pd'^MK^{(g_3)}_{b,\bsk_1}(z')\pd'_NK^{(g_3)}_{c,\bsk_2}(z')\pd'^NE^{(g_3)}_{d}(z') 
\Big] \nn\\
& \!\!\!\!\!\!\!\!\!\!\!\!\! \quad
-\int_{z'} \sum_{\substack{a+b+c+d=n \\ a \geq 0, ~ b,c,d \geq 1}} \!\!\!\!
\pd'_MG_{a,\bsk_1+\bsk_2}(z,z')
\bigg[ 
\frac{1}{2}\pd'^MC^{(g_3)}_{b,\bsk_1,\bsk_2}(z')\pd'_NE^{(g_3)}_{c}(z')\pd'^NE^{(g_3)}_{d}(z') \nn\\[-15pt]
&{} \quad\quad\quad\quad\quad\quad\quad\quad\quad\quad\quad\quad\quad\quad
+\pd'^ME^{(g_3)}_{b}(z')\pd'_NE^{(g_3)}_{c}(z')\pd'^NC^{(g_3)}_{d,\bsk_1,\bsk_2}(z') \bigg] \nn\\
& \!\!\!\!\!\!\!\!\!\!\!\!\! \quad
+\int_{z'} \sum_{\substack{a+b+c=n \\ a \geq 0, ~ b,c \geq 1}} 
G_{a,\bsk_1+\bsk_2}(z,z') \left\{ 
K^{(g_3)}_{b,\bsk_1}(z') K^{(a_2)}_{c,\bsk_2}(z')
+C^{(g_3)}_{b,\bsk_1,\bsk_2}(z') E^{(a_2)}_{c}(z') \right\},
\label{Ca2n}
\end{align}
and the two-point function is derived from $K^{(a_2)}_{n,\bsk_1}$,
\begin{align}
K^{(a_2)}_{n,\bsk_1}(z)
&:= 
\frac{1}{(n-1)!a_2(2g_3)^{n-3}} 
\frac{\de^{n}\Phi^{(a_2)}_{n}(z,\bsk_2)}{\de\vphi_{[0]}(-\bsk_1)\de\vphi_{[0]}(\bsnl)^{n-1}} \nn\\[+5pt]
& \!\!\!\!\!\!\!\!\!\!\!\!\! =
-\int_{z'}  \sum_{\substack{a+b+c+d=n \\ a \geq 0, ~ b,c,d \geq 1}} \!\!\!\!
\pd'_MG_{a,\bsk_1}(z,z')
\Big[ 
\pd'^MK^{(g_3)}_{b,\bsk_1}(z')\pd'_NE^{(g_3)}_{c}(z')\pd'^NE^{(g_3)}_{d}(z') \nn\\[-15pt]
&{} \quad\quad\quad\quad\quad\quad\quad\quad\quad\quad\quad\quad\quad\quad
+2\pd'^ME^{(g_3)}_{b}(z')\pd'_NE^{(g_3)}_{c}(z')\pd'^NK^{(g_3)}_{d,\bsk_1}(z') 
\Big] \nn\\
& \!\!\!\!\!\!\!\!\!\!\!\!\! \quad
+\int_{z'} \sum_{\substack{a+b+c=n \\ a \geq 0, ~ b,c \geq 1}} 
G_{a,\bsk}(z,z')E^{(a_2)}_{b}(z') K^{(g_3)}_{c,\bsk_1}(z'),
\end{align}
where we introduced 
$\pd_M E_n = (\pd_zE_n,\bsnl)$ and $\pd_M \cK_{n,\bsq} = (\pd_z,i\bsq)\cK_{n,\bsq}(z)$, etc. for  momentum space we are working in. Indices are contracted as usual.
We omit the expressions for $E^{(a_2)}_n$ since our results will not involve it at the leading order in $\lm$.

Based on the above results, we can now define the effective propagators, 
\begin{align}
\bsE(z)&:=\sum_{n=1}^\infty (-2g_3)^{n-1}(-\bar\phi)^{n}E^{(g_3)}_{n}(z), \label{bsE} \\
\bsK_{\bsq}(z) 
&:=\sum_{n=1}^\infty(2g_3\bar\phi)^{n-1}K^{(g_3)}_{n,\bsq}(z) \nn\\
&=\sum_{p=0}^\infty(-2g_3)^p\int_{z_1,...,z_p}\cG_{\bsq}(z,z_1)\bsE(z_1)\cG_{\bsq}(z_1,z_2)\cdots\bsE(z_p)\cK_{\bsq}(z_p), \label{bsKq} \\
\wt{\bsE}(z)
&:=\sum_{n=1}^\infty (-2g_3)^{n-3}(-\bar\phi)^{n}a_2 E^{(a_2)}_{n}(z) \nn\\
&=-a_2 \int_{z'} \pd'_M\bsG_{\bsnl}(z,z')\pd'^M\bsE(z')\pd'_N\bsE(z')\pd'^N\bsE(z'), \\
\bsG_{\bsq}(z,z') 
&:= \sum_{n=0}^\infty(2g_3\bar\phi)^n G_{n,\bsq}(z,z') \nn\\
&=\sum_{p=0}^\infty(-2g_3)^p\int_{z_1,...,z_p}\cG_{\bsq}(z,z_1)\bsE(z_1)\cG_{\bsq}(z_1,z_2)\cdots\bsE(z_p)\cG_{\bsq}(z_p,z').
\end{align}
We will only use the bulk-boundary effective propagators $\bsE,\bsK$ explicitly later. Here, we give their diagrammatic representations.

\subsection{Correlation functions at all orders} \label{corr}

We derive AdS$_4$ integral representations of the two-point and three-point functions using the derivatives of the classical solutions found in the last appendix. In what follows, CFT correlation functions in Sec.\,\ref{twopointfunctionB1}, \ref{three-pointB2}, \ref{zeromomentum} and \ref{general-sr-k} are all for CFT dual to AdS, while those defined with CFT dual to dS only appear in Sec.\,\ref{adstods}.
\subsubsection{Two-point function}\label{twopointfunctionB1}
The full two-point function can be written as
\begin{align}
\lan O_0(\bsq)O_0(-\bsq)\ran' 
&=\sum_{n=0}^\infty\frac{(-\bar\phi)^n}{n!}
\lan O_0(\bsq)O_0(-\bsq)O_0(\bsnl)^n \ran'_{\rm CFT} \nn\\
&=
\Bigg[
\sum_{n=0}^\infty(2g_3\bar\phi)^n
K^{(g_3)}_{n+1,\bsq}(z)
+\sum_{n=2}^\infty(-\bar\phi)^{n}(-2g_3)^{n-2}a_2
K^{(a_2)}_{n+1,\bsq}(z)   \Bigg]_{[3-2\lm]}.
\label{two-pointcorrC}
\end{align}
Here the operation $A|_{[3-2\lm]}$ means picking up the coefficient of $z^{3-\lm}$-term in the asymptotic form of $A$ as $z\to 0$ and multiplying it by $(3-2\lm)R_{\rm AdS}^2$ (to put it practically, replacing every first $\cG_{\bsq}(z,z')$ with $\cK_{\bsq}(z')$ in the integrals). Diagrammatically, it means pulling the bulk point with coordinate $z$ to the boundary.
The first term of \eqref{two-pointcorrC} is just the effective bulk-boundary propagator with the bulk point pulled to the boundary,
\begin{align}
\sum_{n=0}^\infty(2g_3\bar\phi)^n
K^{(g_3)}_{n+1,\bsq}(z)\Big|_{[3-2\lm]}
&=
\bsK_{\bsq}(z)\big|_{[3-2\lm]} \nn\\
&\!\!\!\!\!\!\!\!\!\!\!\!\!\!\!\!\!\!\!\!\!\!\!\!\!\!\!\!\!\!\!\!\!\!\!\!\!\!\!=
\sum_{p \geq 0}(-2g_3)^p\int_{z_1,...,z_p} \!\!\!\!\!
\cK_{\bsq}(z_1)\bsE(z_1)\cG_{\bsq}(z_1,z_2)\bsE(z_2)\cdots \cG_{\bsq}(z_{p-1},z_p)\bsE(z_p)
\cK_{\bsq}(z_p),
\end{align}
where $p=0$ in the sum is just the CFT two-point function $\lan O_0(\bsk)O_0(-\bsk) \ran_{\rm CFT}'$,
we find
\begin{align}
\lan O_0(\bsq)O_0(-\bsq) \ran' &{} \nn\\
~~ =\bsK_{\bsq}(z)\big|_{[3-2\lm]}
&{} 
-a_2\int_{z}\pd_M\bsK_{\bsq}(z) \left[ 
\pd^M\bsK_{\bsq}(z)\pd_N\bsE(z)\pd^N\bsE(z) 
+2\pd^M\bsE(z)\pd_N\bsE(z)\pd^N\bsK_{\bsq} (z)
\right] \nn\\
&
+2g_3a_2\int_{z,z'}
\bsK_{\bsq}(z)^2
\pd'_M\bsG_{\bsnl}(z,z')\pd'^M\bsE(z')\pd'_N\bsE(z')\pd'^N\bsE(z') \nn\\
=\bsK_{\bsq}(z)\big|_{[3-2\lm]} 
&{}
-a_2\int dz \left[ 
3\pd_z\bsK_{\bsq}(z)\pd_z\bsK_{\bsq}(z)\pd_z\bsE(z)\pd_z\bsE(z)-\bsq^2 \bsK_{\bsq}(z)^2\pd_z\bsE(z)\pd_z\bsE(z) 
\right] \nn\\
&+\ldots  \,.
\label{full-twopos}
\end{align}
Here we introduced the notation $\pd_M \bsE = (\pd_z\bsE,\bsnl)$ and $\pd_M \bsK_{\bsq} = (\pd_z,i\bsq)\bsK_{\bsq}(z)$. Indices are contracted as usual. We omitted the $\cO(g_3a_2)$ term, which turns out to be subleading in $\lm$.
\subsubsection{Three-point function}\label{three-pointB2}
We begin with the identity
\begin{align}
\lan O_0(\bsk_1)O_0(\bsk_2)O_0(\bsk_3)\ran'
&=
\sum_{n=0}^\infty\frac{(-\bar\phi)^n}{n!}
\lan O_0(\bsk_1)O_0(\bsk_2)O_0(\bsk_3)O_0(\bsnl)^n \ran_{\rm CFT}' \nn\\
&\!\!\!\!\!\!\!\!\!\!\!\!\!\!\!\!\!\!\!\!\!\!\!\!\!\!\!\!\!\!\!\!\!\!\!\!\!\!\!\!\!\!\!\!\!\!\! = 
\Bigg[
\sum_{n=0}^\infty\frac{(-\bar\phi)^n(-2g_3)^{n+1}}{2}
C^{(g_3)}_{n+2,\bsk_2,\bsk_3}(z)
+\sum_{n=1}^\infty(-\bar\phi)^n(-2g_3)^{n-1}a_2 C^{(a_2)}_{n+2,\bsk_2,\bsk_3}(z)
\Bigg]_{[3-2\lm]}.
\label{three-pointcorrC}
\end{align}
Using the recursive relations we found, we can rewrite the three-point function \eqref{three-pointcorrC} as
\begin{align}
&{} 
\lan O_0(\bsk_1)O_0(\bsk_2)O_0(\bsk_3)\ran' \nn\\
&\quad =
-2g_3\int_{z} \bsK_{\bsk_1}(z)\bsK_{\bsk_2}(z)\bsK_{\bsk_3}(z) \nn\\
&{} \quad\quad
-2a_2\int_{z}\pd_M\bsE(z)
\left[
\pd^M\bsK_{\bsk_1}(z)\pd_N\bsK_{\bsk_2}(z)\pd^N\bsK_{\bsk_3}(z)+(231)+(312)
\right] \nn\\
&{} \quad\quad
+(-2g_3)^2\int_{z,z'}
\left[
\wt{\bsE}(z)\bsK_{\bsk_1}(z)\bsG_{\bsk_1}(z,z')\bsK_{\bsk_2}(z')\bsK_{\bsk_3}(z')
+(231)+(312)
\right]
\nn\\
&{} \quad\quad
+2g_3a_2\int_{z,z'}
\Big[
\bsK_{\bsk_1}(z)\bsK_{\bsk_2}(z)
\pd'_M\bsG_{\bsk_3}(z,z') 
\nn\\
&{} \quad\quad\quad\quad\quad\quad\quad\quad\quad\quad\quad
\times\Big( 
\pd'^M\bsK_{\bsk_3}(z')
\pd'_N\bsE(z')
\pd'^N\bsE(z') 
+2\pd'^M\bsE(z')
\pd'_N\bsE(z')
\pd'^N\bsK_{\bsk_3}(z')
\Big) 
\nn\\
&{} \quad\quad\quad\quad\quad\quad\quad\quad\quad~~
+(231)+(312)\Big]
\label{three-pointfull} \\
&\quad =
-2g_3\int \frac{dz}{z^4} \bsK_{\bsk_1}(z)\bsK_{\bsk_2}(z)\bsK_{\bsk_3}(z)  \label{three-pointfull-mom1} \\
&{} \quad\quad
-2a_2\int{dz}
\Big[3\pd_z \bsK_{\bsk_1}(z)\pd_z\bsK_{\bsk_2}(z)\pd_z\bsK_{\bsk_3}(z)\pd_z\bsE(z)
\nn\\
& \quad\quad\quad\quad\quad\quad\quad\quad~~
-\Big(\boldsymbol k_1.\boldsymbol k_2 {\bsK}_{\bsk_1}(z){\bsK}_{\bsk_2}(z)\pd_z\bsK_{\bsk_3}(z)\pd_z\bsE(z) + (231)+(312)\Big) \Big] 
\label{three-pointfull-mom2} \\
&\quad\quad + \cdots\,. \nn
\end{align}
Here we dropped the other integrals because they turn out to be subleading in $\lm$ and so only the first and second integrals \eqref{three-pointfull-mom1}, \eqref{three-pointfull-mom2} contribute to the three-point function. The first integral \eqref{three-pointfull-mom1} comes from \eqref{Cg3n} while the second comes from the first integral of \eqref{Ca2n}.

\section{Explicit form of the correlation functions at all orders} \label{explicit}
In this appendix, we evaluate the effective bulk-boundary propagators $\bsE$ and $\bsK_\bsq$ explicitly at the leading order in $\lm$. The non-derivative part of the three-point function will also be evaluated.
\subsection{Zero-momentum bulk-boundary effective propagator}\label{zeromomentum}
Let us compute $E^{(g_3)}_{n}(z)$ at the leading order in $\lm$. We will encounter  the following integral
\begin{align}
\int_0^\infty\frac{dz'}{z'^4}z'^{a\lm}z'^{b\lm}\cG_\bsnl(z,z'),
\end{align}
where the zero-momentum bulk-bulk propagator reads
\begin{align}
\cG_\bsnl(z,z')=\frac{1}{3-2\lm}\left[z^\lm z'^{3-\lm}\te(z-z')+z'^\lm z^{3-\lm}\te(z'-z)\right].
\end{align}
By the step functions in $\cG_\bsnl$, the integral splits into $\int_0^z$ and $\int_z^\infty$ parts. The $\int_z^\infty$ part has an integrand $z'^{(a+b+1)\lm-4}$, so that the integral is convergent around $z\sim\infty$. Also it is regular at $\lm=0$. On the other hand, $\int_0^z$ part has integrand $z'^{-1+(a+b-1)\lm}$ and the leading order of the integral is of $\cO(\lm^{-1})$. Thus the integral becomes
\begin{align}
\int_0^\infty\frac{dz'}{z'^4}z'^{a\lm}z'^{b\lm}\cG_0(z,z')=\frac{z^{(a+b)\lm}}{3(a+b-1)\lm}+\ldots\,,
\label{zeroeffpropint}
\end{align}
where the dots stand for higher order terms in $\lambda$. Here one would wonder if we may approximate that $z^{(a+b)\lm}\simeq1$ at the leading order in $\lambda$. However, this approximation breaks down when $\lm\ln|z|\gtrsim1$ (indeed, it happens for very small $z$ in the integral region $[0,\infty)$). We therefore keep $z^{(a+b)\lm}$ without reducing it to $1$.
This remark applies to the computations in what follows.
With the explicit forms for $n=1$ and $n=2$
\begin{align}
E^{(g_3)}_1(z) &= z^\lambda, \quad\quad
E^{(g_3)}_2(z) = \frac{z^{2\lambda}}{6\lambda},
\end{align}
the recursive definition of $E^{(g_3)}_n$ given by \eqref{Eg1} and \eqref{Egn} yields
\begin{align}
E^{(g_3)}_{n}(z)
=\frac{z^{n\lm}}{(6\lm)^{n-1}}.
\label{E-gen}
\end{align}

\paragraph{Order estimate of the integral \eqref{three-pointfull} for three-point function}
Using this explicit form \eqref{E-gen}, we can estimate the $\lm$-order of the integral \eqref{three-pointfull} for the three-point function to show that \eqref{three-pointfull-mom1} and \eqref{three-pointfull-mom2} dominate.\footnote{The following order estimate can be justified recursively.} 
The point is that one $z$-differentiation of the effective zero-momentum propagator $E^{(g_3)}_{n+1}(z)$ yields one $\lambda$. By this property, we can see for each order in $\bar\phi$, or equivalently in $g_3$, that the 1st and 2nd term on the right hand side of \eqref{three-pointfull} always dominate over the 3rd and 4th terms at the leading order in $\lm$.

\subsection{Effective bulk-boundary propagators  of general momentum}\label{general-sr-k}
Let us compute $K_{n,\bsq}(z)$. We start from $K_{1,\bsq}(z)=\cK_\bsq(z)$.
Applying the recursion relation \eqref{Kgn} to this, we find
\begin{align}
{K_{2,\bsq}(z)}&=
\int_0^\infty\frac{dz'}{z'^4}
\cG_\bsq(z,z'){E^{(g_3)}_1(z')}{\cK_\bsq(z')}.
\label{rec-K1}
\end{align}
The bulk-to-boundary propagator may be expanded as
\begin{align}
\mathcal{K}_\bsq(z)
=q^{-\lm}\left\{
(qz)^\lambda \left[1
+\mathcal{O}\left((qz)^2\right)\right]
+\frac{1}{3}(qz)^{3-\lambda}\Big[1+\mathcal{O}\left((qz)^2\right)\Big]
\right\}\,.
\end{align}
Note that the series expansion is in the the dimensionless combination $qz$. Here and in what follows, we will recurrently use such a dimensionless combination. It will make the computations simpler and also be useful to reproduce the correct scale-dependence of the correlation functions.

The $z>z'$ part of the bulk-to-bulk propagator may be expanded as
\begin{align}
\cG_{\bsq} (z,z')
&=\mathcal{K}_\bsq(z)\mathcal{I}_\bsq(z')
=q^{-3+\lm}\mathcal{K}_\bsq(z)\frac{\Gamma(3/2-\lambda)}{2\Gamma(5/2-\lambda)}
(qz)'^{3-\lambda}\Big[1+\mathcal{O}\left((qz')^2\right)\Big]\,,
\end{align}
where we introduced
\begin{align}
\mathcal{I}_\bsq(z)=2^{1/2-\lambda}\Gamma(3/2-\lambda)(qz)^{3/2}q^{-3+\lambda}I_{3/2-\lambda}(qz)\,.
\end{align}
For $\lambda\ll1$, we may simplify it as
\begin{align}
\cG_{\bsq} (z,z')&\simeq\frac{1}{3}q^{-3+\lm}\mathcal{K}_\bsq(z)(qz')^{3-\lambda}\Big[1+\mathcal{O}\left((qz')^2\right)\Big]\,.
\end{align}
Note that we ignored $\lm$ in the Gamma functions but kept $(qz')^{3-\lambda}$. This is because the approximation $(qz)^\lambda\simeq1$ is not valid when $\lambda|\ln (qz)|\gtrsim 1$, as explained in the zero-momentum case. 
On the other hand, the $z'>z$ part may be expanded as
\begin{align}
\cG_{\bsq} (z,z')
&=q^{-\lm}\mathcal{I}_\bsq(z)\left[(qz')^\lambda \left(1
+\mathcal{O}\left((qz')^2\right)\right)
+\frac{1}{3}(qz')^{3-\lambda}\Big(1+\mathcal{O}\left((qz')^2\right)\Big)\right].
\end{align}
The recursion relation for $n=2$ \eqref{rec-K1} reads, reflecting the step functions in $\cG_\bsq$,
\begin{align}
K_{2,\bsq}(z) &= \left(\int_0^z+\int_z^\infty\right) \frac{dz'}{z'^4} \mathcal G_{\bsq} (z,z')
E^{(g_3)}_1(z') \mathcal{K}_{\bsq}(z') \,.
\label{rec-n2-2}
\end{align}
The first part is
\begin{align}
\int_0^z \mbox{part }
\simeq ~ q^{-\lm}\mathcal{K}_\bsq(z) \int_0^{qz}\frac{dy'}{y'^4} \frac{1}{3}y'^{3+\lambda}\Big[1+\mathcal{O}\left(y'^2\right)\Big]\,, \quad y'=qz',
\label{0z}
\end{align}
whose dominant contribution is from $y'\sim0$. We can find the leading order terms in $\lm$ by collecting the $y'^{-1+\cO(\lm)}$ terms in the integrand. The remaining part of \eqref{rec-n2-2} is
\begin{align}
\int_z^\infty \mbox{part }
\simeq ~ q^{3-3\lm}\mathcal{I}_q(z)\left(\int_0^\infty\!\!-\int_0^{qz}\right)\frac{dy'}{y'^4}y'^\lambda 
\left[y'^\lambda \left(1
+\mathcal{O}\left(y'^2\right)\right)
+\frac{1}{3}y'^{3-\lambda}\left(1+\mathcal{O}\left(y'^2\right)\right)\right]^2\,.
\label{0infty-0z}
\end{align}
Its dominant contribution is again from $y'\sim0$ since the integrand is originally proportional to $y'^{\lm-4}\cK_\bsq^2(y')$, which damps exponentially as $y'\to\infty$. Therefore the leading order terms in $\lm$ can be found by collecting the $y'^{-1+\cO(\lm)}$ terms and applying \eqref{xaexp} to them. Then it turns out that  the contributions from $\int_0^\infty$ and $\int_0^{qz}$ differ by an overall factor of $(qz)^\lambda$, which is not negligible when $\lambda |\ln (qz)|\gtrsim1$. Whether it is negligible  is determined by whether the $(qz)$-integral to which the effective propagator is applied is regular around $qz=0$ or not. We will discuss this point for the integrals \eqref{three-pointfull-mom1} and \eqref{three-pointfull-mom2}, respectively.

\subsubsection{Effective propagators in integral \eqref{three-pointfull-mom1}}\label{slowroll}
Let us consider the case in which the effective propagators are used in the integral \eqref{three-pointfull-mom1}. We first estimate the order in $\lambda$. For this, following the technique given in Appendix \ref{tripleKintegrals}, we collect terms of the form $z^{-1+\cO(\lambda)}$, which gives rise to $\lambda^{-1}$. Indeed the integral \eqref{three-pointfull-mom1} contains $z^{-1+\cO(\lambda)}$. So the $\lambda$-expansion of this integral starts from order $\lambda^{-1}$. Since this integrand  $z^{-1+\cO(\lambda)}$ is singular around $z=0$, it is dangerous to set $(qz)^\lm=1$ and thus \eqref{0infty-0z} cannot be zero.
 
To extract the leading order in $\lm$ from the integral \eqref{rec-n2-2}, namely \eqref{0z}$+$\eqref{0infty-0z}, we collect terms of the form $y'^{-1+\cO(\lm)}$ following Appendix \ref{tripleKintegrals}. Then we find
\begin{align}\label{K1k}
K_{2,\bsq}(z) 
& = q^{-2\lm}\lm^{-1} \left[
\frac{2}{9} (qz)^{3-\lambda} -\frac{1}{9} (qz)^3 +\frac{1}{3}(qz)^{2\lambda} + \mbox{higher order in} ~ \lm \right].
\end{align}
Using the recursion relation \eqref{rec-Kn}, we can find the structure of $K_{n+1,\bsq}(z)$ inductively
\begin{align}
K_{n+1,\bsq}(z) = q^{-(n+1)\lm}\lm^{-n}
\bigg[A_{n}(qz)^{(n+1)\lm}+\sum_{l=0}^{n}B_{nl}(qz)^{3+(l-1)\lm}\bigg],
\label{Kn1k-SR}
\end{align}
where $A_n$, $B_{nl}$ are independent of $\lm$. 

Let us evaluate the integral \eqref{three-pointfull-mom1}. This involves three momenta $k_1,k_2,k_3$. We introduce $k=k_1+k_2+k_3$ as a reference momentum and rescale the integration variable as $y=kz$. (See also Appendix \ref{tripleKintegrals}.) We can then use the approximation $(k_i/k)^\lm\sim 1$ by the approximation \eqref{approximation}. Then, we can show at the leading order in $\lm$ that the integral is proportional to $k_1^{3-3\lambda}+k_2^{3-3\lambda}+k_3^{3-3\lambda}$ at all orders in $\bar{\phi}$. The proportional factor is the function of $\bar\phi k^{-\lm}$, which is fixed in Sec.\,\ref{RG}.

\subsubsection{Effective propagators in integral \eqref{three-pointfull-mom2}}\label{kmomentum}
Let us next consider the case in which the effective propagators are used in the integral \eqref{three-pointfull-mom2}. We first estimate the order in $\lambda$. In this case it turns out that the leading contribution in the $\lm$-expansion comes from the terms in the integrand which are regular around $z=0$. So we may set $(qz)^\lm=1$ and thus \eqref{0infty-0z} becomes zero. Thus it is enough to evaluate the $[0,z]$-part \eqref{0z}.
Then, $K_{2,\bsq}(z)$ becomes
\begin{align}
K_{2,\bsq}(z) &= \frac{z^\lambda}{3\lambda} \mathcal K_{\bsq} (z).
\end{align}
For $K_{n,\bsq}(z)$ with general $n$, it is also enough to evaluate the $z'$-integral over $[0,z]$ in the recursion relation \eqref{rec-Kn}.
We then find
\begin{align}\label{Knk}
K_{n+1,\bsq}(z)= (n+1)\left( \frac{z^\lambda}{6\lambda} \right)^{n} \mathcal K_{\bsq}(z).
\end{align}

Let us find explicit forms of $\bsK_\bsq(z)$ and $\bsE(z)$. Here $\bsq$ is one of the three-momenta $\bsk_1,\bsk_2,\bsk_3$ in the three-point function of our interest. 
Before taking the infinite sum in the definitions \eqref{bsE} and \eqref{bsKq} of $\bsE$ and $\bsK_\bsq$, let us clarify the power structure of $\bsK_\bsq(z)$ and $\bsE(z)$. As in the non-derivative case, we rescale the integration variable with the reference momentum $k$ as $y=kz$ in \eqref{three-pointfull-mom1} and \eqref{three-pointfull-mom2}. Then we can see that $-\bar\phi^{-1}\bsE$ and $\bsK_\bsq$ are the power series in the following form: for each $n$
\[
(-\bar\phi)^{n}E^{(g_3)}_{n}, ~~ (-\bar\phi)^{n-1}K^{(g_3)}_{n,\bsq} 
\quad\propto\quad \left(\frac{-\bar\phi k^{-\lm}}{\lm}\right)^n(1+\cO(\lm)),
\] 
where the approximation \eqref{approximation} was applied. Therefore, in the IR scale characterized by $\bar\phi k^{-\lm}/\lm=\cO(1)$, it is enough to pick up the $\lm$-leading contributions to $K_{n,\bsq}$ and $-\bar\phi^{-1}E^{(g_3)}_{n}$ for each $n$ in order to extract the $\lm$-leading contributions to \eqref{bsE} and \eqref{bsKq}.
Substituting \eqref{E-gen} and \eqref{Knk} into the definitions \eqref{bsE} and \eqref{bsKq}, we find
\begin{align}
\bsK_{\bsq}(z) 
&= \left(1-\frac{g_3 \bar \phi z^\lambda}{3\lambda} \right)^{-2}\mathcal K_{\bsq}(z), \\
\bsE(z) 
&= (-\bar\phi)\left( 1-\frac{g_3 \bar \phi z^\lambda}{3\lambda} \right)^{-1} z^\lm.
\end{align}
Since we can use the approximation $(kz)^\lm=1$ inside the integral \eqref{three-pointfull-mom2}, these forms can be simplified further,
\begin{align}
\bsK_{\bsq}(z) 
&= \left(1-\frac{g_3 \bar \phi k^{-\lambda}}{3\lambda} \right)^{-2}\mathcal K_{\bsq}(z), \\
\bsE(z) 
&= (-\bar\phi)\left( 1-\frac{g_3 \bar \phi k^{-\lambda}}{3\lambda} \right)^{-1} z^\lm.
\end{align}
Similarly, we may simplify $\bsK'_{\bsk_i}(z) $ and $\bsE'(z) $ as
\begin{align}
\label{K'AdS}
\bsK'_{\bsq}(z) 
&= {\bigg( 1-\frac{g_3 \bar \phi k^{-\lambda}} {3\lambda} \bigg)^{-2} } \mathcal K'_{\bsq}(z)\,, \\
\label{E'AdS}
\bsE'(z) 
&= (-\lambda\bar\phi){\bigg( 1-\frac{g_3 \bar \phi k^{-\lambda}} {3\lambda} \bigg)^{-2} } z^{\lambda-1}\,.
\end{align}

\subsection{From AdS to dS} \label{adstods}
We have obtained from the AdS action \eqref{AdSmodel} the integral expressions for the full two-point  and three-point functions and the explicit forms of the effective propagators. Now, following Sec.\,\ref{UVCFT}, we convert them into the forms associated with the UV CFT dual to our inflation model (in dS). To achieve this, we apply the replacement \eqref{Mflip} after recovering the AdS radius dependence using the rule \eqref{rads}. The result is the following: \\
(i) effective bulk-boundary propagators for \eqref{three-pointfull-mom2}
\begin{align}
\bsK^{\rm dS}_{\bsq}(z)
&= \left(1+\frac{g_3 R_{\rm dS}^{-2} \bar \phi k^{-\lambda}}{3\lambda} \right)^{-2}\mathcal K_{\bsq}(z) , 
\label{eftKdS} \\
\bsE^{\rm dS}(z) 
&= -\bar\phi\left( 1+\frac{g_3 R_{\rm dS}^{-2} \bar \phi k^{-\lambda}}{3\lambda} \right)^{-1} z^\lm. \label{eftEdS}
\end{align}
Their derivatives corresponding to~\eqref{K'AdS} and~\eqref{E'AdS} may also be obtained in a similar way.
\\
(ii) integral form of the two-point function
\begin{align}
&{}
\lan O_0(\bsq)O_0(-\bsq) \ran' \nn\\
&~~ =-R_{\rm dS}^2\bsK_{\bsq}(z)\big|_{[3-2\lm]}
 \nn\\
&{} \quad\quad
+\sum_{m \geq 2}\frac{2m}{2^m}\al_m (R_{\rm dS})^{2(m-2)} 
\int_0^\infty dz z^{2(m-2)} \bsE'(z)^{2(m-1)}
\left\{ 
(2m-1)\bsK_{\bsq}'(z)^2-\bsq^2 \bsK_{\bsq}(z)^2 
\right\}. \label{full-twopos-dS}
\end{align}
(iii) integral form of the three-point function
\begin{align}
&{} 
\lan O_0(\bsk_1)O_0(\bsk_2)O_0(\bsk_3)\ran' \nn\\
&~~ =
-2g_3\int_{z} \bsK_{\bsk_1}(z)\bsK_{\bsk_2}(z)\bsK_{\bsk_3}(z) \nn\\
&{} \quad~~
+\sum_{m \geq 2}\frac{m(m-1)}{2^{m-2}}\al_m(R_{\rm dS})^{2(m-2)} \!\!
\int_0^\infty \!\! {dz} z^{2(m-2)}{\bsE}'(z)^{2m-3}
\Big\{(2m-1)
{\bsK}_{\bsk_1}'(z){\bsK}_{\bsk_2}'(z){\bsK}_{\bsk_3}'(z)
\nn\\
& \quad\quad\quad\quad\quad\quad\quad\quad\quad
- \Big(\boldsymbol k_1.\boldsymbol k_2 {\bsK}_{\bsk_1}(z){\bsK}_{\bsk_2}(z){\bsK}_{\bsk_3}'(z) + (231)+(312)\Big) \Big\}.
\label{full-threepos-dS}
\end{align}

\section{three-point functions with two nearly marginal scalars}
\label{tripleKintegrals}
In this appendix we compute CFT three-point functions
\begin{align}
\lan O_1(\bsk_1)O_2(\bsk_2)O_3(\bsk_3) \ran'_{\rm CFT}\,
\end{align}
of  two nearly marginal scalars, $O_1$ and $O_2$, and one scalar $O_3$ with a conformal dimension close to a non-negative integer $n$. More explicitly, we parametrize the dimension $\Delta_i$ of $O_i$ as
\begin{align}
\label{two_marginal}
\Delta_1=3+v_1\lambda\,,
\quad
\Delta_2=3+v_2\lambda\,,
\quad
\Delta_3=n+v_3\lambda\,,
\end{align}
where $\lambda\ll1$ characterizes the deviation from marginality and integer dimension, and we focus on the leading order in $\lambda$ of the three-point functions.

\medskip
As we mentioned earlier, a CFT three-point function of primary scalars of dimensions $\De_i~(i=1,2,3)$ can be uniquely fixed by the conformal symmetry as~\cite{Bzowski:2015pba}
\begin{align}
&\lan O_1(\bsk_1)O_2(\bsk_2)O_3(\bsk_3) \ran' = C_{123} k^{\Delta_1+\Delta_2+\Delta_3-3} I_{\frac{d}{2}-1\{\De_1-\frac{3}{2},\De_2-\frac{3}{2},\De_3-\frac{3}{2}\}}(\kappa_1,\kappa_2,\kappa_3)
\end{align}
up to the OPE coefficient $C_{123}$. Here we introduced $k=k_1+k_2+k_3$ and $\kappa_i=k_i/k$. 
We refer to the function $I_{\alpha\{\beta_1,\beta_2,\beta_3 \}}$ as the triple-$K$ integral \cite{Bzowski:2013sza,Bzowski:2015pba, Bzowski:2015yxv} defined by\footnote{Note that the triple-$K$ integrals take the same form, up to an overall numerical constant, as the three-point correlation function computed with the Witten diagrams in $(d+1)$-dimensional AdS spacetime, in which the coefficient $C_{123}$ is determined from the bulk action in AdS$_{d+1}$, and the $x$-integral in \eqref{3kint} corresponds to the integral in the radial direction of AdS$_{d+1}$.}
\begin{align}
I_{\alpha\{\beta_1,\beta_2,\beta_3 \}}(\kappa_1,\kappa_2,\kappa_3) = \kappa_1^{\beta_1}\kappa_2^{\beta_2}\kappa_3^{\beta_3} \int_0^\infty dx x^\alpha K_{\beta_1}(\kappa_1 x) K_{\beta_2}(\kappa_2 x)K_{\beta_3}(\kappa_3 x)\,,
\label{3kint}
\end{align}
where $K_\nu(z)$ is the modified Bessel function of the second kind.
Refs. \cite{Bzowski:2015pba, Bzowski:2015yxv} studied properties of the triple-$K$ integral in detail. Here we give a brief summary of their results we will use. The integral \eqref{3kint} converges if the parameters satisfy $\al+1>|\bt_1|+|\bt_2|+|\bt_3|$ for fixed $\kappa_1,\kappa_2,\kappa_3>0$. Outside this parameter region, we define the integral as its maximal analytical continuation such that the integral coincides with the integral evaluated in the convergence region. This integral, however, still has singularities for special parameter sets, defined by $\al+1\pm\bt_1\pm\bt_2\pm\bt_3=-2n$ for $n=0,1,...$, where the choice of the three $\pm$ is arbitrary. In particular, for our parameter set~\eqref{two_marginal}, the triple-$K$ integral is singular in the limit $\lambda\to0$. In the following we compute the triple-$K$ integral for~\eqref{two_marginal} at the leading order in $\lambda$ by applying the discussion in~\cite{Bzowski:2015yxv}.

\subsection{General results}
By using the series expansion of the Bessel $K$ function,
\begin{align}
K_{\nu}(z)=\sum_{j=0}^\infty \left[ a_j^-(\nu)z^{-\nu+2j}+a_j^+(\nu)z^{\nu+2j} \right] \quad{\rm with}\quad
a_j^\pm(\nu):=\frac{(-)^j\Ga(\mp\nu-j)}{2^{\pm\nu+2j+1}j!}\,,
\label{Kexp}
\end{align}
we notice that the integrand of the triple-$K$ integral of our interest contains terms schematically of the form,
\begin{align}
\sim \sum_{j=0}^\infty x^{-1\pm n+2j+\mathcal{O}(\lambda)}\,,\quad
\sum_{j=0}^\infty x^{-1\pm (n-3)+2j+\mathcal{O}(\lambda)}\,,
\quad
\sum_{j=0}^\infty x^{-1-(n-6)+2j+\mathcal{O}(\lambda)}\,.
\end{align}
Here for notational simplicity we omitted the coefficients of each term, which are  of $\mathcal{O}(\lambda^0)$. An important point is that there always appear $x^{-1+\cO(\lm)}$-terms when $n\geq0$, i.e., when $O_3$ has a non-negative dimension close to an integer. Since $ \int dx\, x^{-1+\alpha\lambda}= \frac{1}{\alpha\lambda} x^{\alpha\lambda}$, one may naively expect that the singularity of the triple-$K$ integral in the $\lambda\to0$ limit is associated with those $x^{-1+\cO(\lm)}$-terms. Indeed, Ref.~\cite{Bzowski:2015yxv} showed that the singularity in this limit arises only from the integration over a region around $x=0$ of $x^{-1+\cO(\lm)}$-terms, by virtue of the maximal analytic continuation. Based on the argument in~\cite{Bzowski:2015yxv}, we apply the following recipe to compute the triple-$K$ integral at the leading order in $\lambda$:
\begin{enumerate}
\item
Expand the integrand of the triple-$K$ integral \eqref{3kint} in $x$ using~\eqref{Kexp}.
\item Then, collect the $x^{-1+\cO(\lm)}$-terms and integrate them over a region around $x=0$ using
\begin{align}
\int_0^{\mu}dx\,x^{-1+\alpha\lambda} 
= \frac{1}{\alpha\lambda}\mu^{\alpha\lambda}
= \frac{1}{\alpha\lambda}\left[1+\mathcal O(\alpha\lambda\ln\mu)\right]\,,
\label{xaexp}
\end{align}
where $\mu\ll 1$ is the cutoff satisfying the condition $|\alpha\lambda\ln\mu|\ll1$.

\end{enumerate}
This procedure gives us the leading contribution to the triple-$K$ integrals in the $\lm$ expansion.

\medskip
Now it is straightforward to compute the triple-K integral. Below, we first explain the computation for the $n=0$ case in detail, and then present the results for general $n$.

\paragraph{Detailed explanation for $\bf n=0$:}

We start from a detailed explanation for the $n=0$ case. In this case, the integrand has three types of $x^{-1+\cO(\lm)}$-terms:
\begin{align}
\nonumber
&x^{-1-(v_1+v_2+v_3)\lambda}a_{0}^{-}(\tfrac{3}{2}+v_1\lambda)\,a_{0}^{-}(\tfrac{3}{2}+v_2\lambda)a_{0}^{-}(-\tfrac{3}{2}+v_3\lambda)
\\
\nonumber
&+(\kappa_1)^{3+2v_1\lambda}(\kappa_3)^{-3+2v_3\lambda}x^{-1+(v_1-v_2+v_3)\lambda}\,a_{0}^{+}(\tfrac{3}{2}+v_1\lambda)a_{0}^{-}(\tfrac{3}{2}+v_2\lambda)a_{0}^{+}(-\tfrac{3}{2}+v_3\lambda)
\\
&+(\kappa_2)^{3+2v_2\lambda}(\kappa_3)^{-3+2v_3\lambda}x^{-1+(-v_1+v_2+v_3)\lambda}\,a_{0}^{-}(\tfrac{3}{2}+v_1\lambda)a_{0}^{+}(\tfrac{3}{2}+v_2\lambda)a_{0}^{+}(-\tfrac{3}{2}+v_3\lambda)\,.
\end{align}
By applying Eq.~\eqref{xaexp}, we find
\begin{align}
\nonumber
&I_{\frac{1}{2}\{\frac{3}{2}+v_1\lm,\frac{3}{2}+v_2\lm,-\frac{3}{2}+v_3\lm\}}(k_1,k_2,k_3)
\\
\nonumber
&=\frac{a_{0}^{-}(\tfrac{3}{2})\,a_{0}^{-}(\tfrac{3}{2})a_{0}^{-}(-\tfrac{3}{2})}{-(v_1+v_2+v_3)\lambda}
+\left(\frac{\kappa_1}{\kappa_3}\right)^3\frac{a_{0}^{+}(\tfrac{3}{2})\,a_{0}^{-}(\tfrac{3}{2})a_{0}^{+}(-\tfrac{3}{2})}{(v_1-v_2+v_3)\lambda}
+\left(\frac{\kappa_2}{\kappa_3}\right)^3\frac{a_{0}^{-}(\tfrac{3}{2})\,a_{0}^{+}(\tfrac{3}{2})a_{0}^{+}(-\tfrac{3}{2})}{(-v_1+v_2+v_3)\lambda}+\mathcal{O}(\lambda^0)
\\
\label{n=0}
&=\frac{\pi^{3/2}}{6\sqrt{2}\,\lambda}\left[
\frac{1}{-(v_1+v_2+v_3)}
+\frac{(\kappa_1/\kappa_3)^3}{v_1-v_2+v_3}
+\frac{(\kappa_2/\kappa_3)^3}{-v_1+v_2+v_3}
\right]+\mathcal{O}(\lambda^0)\,.
\end{align}
Here we have used $(\kappa_i)^\lambda=1+\lambda \ln \kappa_i+\ldots=1+\mathcal{O}(\lambda)$. This approximation is applicable as long as $|\lambda\ln (k_i/k_t)|\ll1$, which is indeed the case \eqref{approximation} in our inflationary discussions.

\paragraph{three-point functions for $\bf n=1,2,3$:}

In a similar manner we may compute the integral for $n=1,2,3$ as follows: For $n=1$,
\begin{align}
I_{\frac{1}{2}\{\frac{3}{2}+v_1\lm,\frac{3}{2}+v_2\lm,-\frac{1}{2}+v_3\lm\}}(\kappa_1,\kappa_2,\kappa_3) =\frac{\pi^{3/2}}{4\sqrt{2}(v_1+v_2-v_3)\lambda}\frac{(\kappa_1)^2+(\kappa_2)^2-(\kappa_3)^2}{\kappa_3}+\mathcal{O}(\lambda^0)\,.
\end{align}
For $n=2$,
\begin{align}
I_{\frac{1}{2}\{\frac{3}{2}+v_1\lm,\frac{3}{2}+v_2\lm,\frac{1}{2}+v_3\lm\}}(\kappa_1,\kappa_2,\kappa_3) =\frac{\pi^{3/2}}{4\sqrt{2}(v_1+v_2+v_3)\lambda}\Big[(\kappa_1)^2+(\kappa_2)^2-(\kappa_3)^2\Big]+\mathcal{O}(\lambda^0)\,.
\end{align}
For $n=3$,
\begin{align}
\label{alphaone}
&I_{\frac{1}{2}\{\frac{3}{2}+v_1\lm,\frac{3}{2}+v_2\lm,\frac{3}{2}+v_3\lm\}}(\kappa_1,\kappa_2,\kappa_3) \nn\\
&= \frac{\pi^{3/2}}{6\sqrt{2}\lambda}
\Bigg[
\frac{(\kappa_3)^{3}}{-v_1-v_2+v_3}
+\frac{(\kappa_2)^{3}}{-v_1+v_2-v_3}
+\frac{(\kappa_1)^{3}}
{v_1-v_2-v_3}
\Bigg] 
+O(\lm^0)\,.
\end{align}
In particular, when $v_1=v_2=v_3=-1$, i.e., $\Delta_1=\Delta_2=\Delta_3=3-\lambda$, it reduces to the form,
\begin{align}
I_{\frac{1}{2}\{\frac{3}{2}-\lm,\frac{3}{2}-\lm,\frac{3}{2}-\lm\}}(\kappa_1,\kappa_2,\kappa_3) 
&= \frac{\pi^{3/2}}{6\sqrt{2}\lambda}
\Big[
(\kappa_1)^{3}
+(\kappa_2)^{3}
+(\kappa_3)^{3}
\Big] 
+O(\lm^0)\,.
\end{align}

\paragraph{General expression for $\bf n\geq4$:}

It is not difficult to derive the general expression for $n\geq4$. For an even $n\geq4$, we have 
\begin{align}
&
I_{\frac{1}{2}\{\frac{3}{2}+v_1\lm,\frac{3}{2}+v_2\lm,n-\frac{3}{2}+v_3\lm\}}(\kappa_1,\kappa_2,\kappa_3) \nn\\
&=
-\frac{(-)^{\frac{n}{2}}}{2^{\frac{3}{2}}(v_1+v_2+v_3)\lm}
\sum_{\substack{j_1+j_2+j_3=\frac{n}{2} \\ j_1,j_2,j_3\geq 0}} \!\!\!\!\!\!
\frac{(\kappa_1)^{2j_1}(\kappa_2)^{2j_2}(\kappa_3)^{2j_3}}{j_1!j_2!j_3!}
{\Ga\left(\tfrac{3}{2}-j_1\right)\Ga\left(\tfrac{3}{2}-j_2\right)\Ga\left(n-\tfrac{3}{2}-j_3\right)} \nn\\
&\quad
-\frac{(-)^{\frac{n}{2}}(\kappa_1)^{3}(\kappa_2)^{3}}{2^{\frac{3}{2}}(v_1+v_2-v_3)\lm}
\sum_{\substack{j_1+j_2+j_3=\frac{n-6}{2} \\ j_1,j_2,j_3\geq 0}} \!\!\!\!\!\!
\frac{(\kappa_1)^{2j_1}(\kappa_2)^{2j_2}(\kappa_3)^{2j_3}}{j_1!j_2!j_3!}
{\Ga\left(-\tfrac{3}{2}-j_1\right)\Ga\left(-\tfrac{3}{2}-j_2\right)\Ga\left(n-\tfrac{3}{2}-j_3\right)} \nn\\
 &\quad+ O(\lm^0)\,,
\end{align}
where the second term should be omitted when $n=4$. Similarly, a general expression for an odd $n\geq5$ is
\begin{align}
&
I_{\frac{1}{2}+u\lm\{\frac{3}{2}+v_1\lm,\frac{3}{2}+v_2\lm,n-\frac{3}{2}+v_3\lm\}}(\kappa_1,\kappa_2,\kappa_3) \nn\\
&=
\frac{(-)^{\frac{n+3}{2}}(\kappa_2)^{3}}{2^{\frac{3}{2}}(-v_1+v_2-v_3)\lm}
\sum_{\substack{j_1+j_2+j_3=\frac{n-3}{2} \\ j_1,j_2,j_3\geq 0}}
\frac{(\kappa_1)^{2j_1}(\kappa_2)^{2j_2}(\kappa_3)^{2j_3}}{j_1!j_2!j_3!}
{\Ga\left(\tfrac{3}{2}-j_1\right)\Ga\left(-\tfrac{3}{2}-j_2\right)\Ga\left(n-\tfrac{3}{2}-j_3\right)}
\nn\\
&+
\frac{(-)^{\frac{n+3}{2}}(\kappa_1)^{3}}{2^{\frac{3}{2}}(v_1-v_2-v_3)\lm}
\sum_{\substack{j_1+j_2+j_3=\frac{n-3}{2} \\ j_1,j_2,j_3\geq 0}}
\frac{(\kappa_1)^{2j_1}(\kappa_2)^{2j_2}(\kappa_3)^{2j_3}}{j_1!j_2!j_3!}
{\Ga\left(-\tfrac{3}{2}-j_1\right)\Ga\left(\tfrac{3}{2}-j_2\right)\Ga\left(n-\tfrac{3}{2}-j_3\right)} \nn\\
&+ O(\lm^0).
\end{align}

\subsection{Squeezed limit}
We then discuss the squeezed limit $\kappa_1\ll\kappa_2=\kappa_3=\kappa$ of the triple-$K$ integral obtained in the previous subsection. For $n=0,1,2,3$, we may use the concrete expressions~\eqref{n=0}-\eqref{alphaone} to conclude that the triple-$K$ integral for $n=1,2$ vanishes in the squeezed limit for arbitrary $v_i$'s, whereas that for $n=0,3$, it does not vanish for a generic set of $v_i$'s. For an even $n\geq4$, the squeezed limit is
\begin{align}
\nonumber
&\lim_{\kappa_1\to0}I_{\frac{1}{2}\{\frac{3}{2}+v_1\lm,\frac{3}{2}+v_2\lm,n-\frac{3}{2}+v_3\lm\}}(\kappa_1,\kappa,\kappa)
\\
\label{squeezed_even}
&=-\frac{(-)^{\frac{n}{2}}\Ga\left(\tfrac{3}{2}\right)\kappa^{n}}{2^{\frac{3}{2}}(v_1+v_2+v_3)\lm}
\sum_{\substack{j_2+j_3=\frac{n}{2} \\ j_2,j_3\geq 0}} \!\!\!\!\!\!
\frac{\Ga\left(\tfrac{3}{2}-j_2\right)\Ga\left(n-\tfrac{3}{2}-j_3\right)}{j_2!j_3!}
+ O(\lm^0)\,.
\end{align}
Similarly, for an odd $n\geq5$, we have
\begin{align}
&
\lim_{\kappa_1\to0}I_{\frac{1}{2}+u\lm\{\frac{3}{2}+v_1\lm,\frac{3}{2}+v_2\lm,n-\frac{3}{2}+v_3\lm\}}(\kappa_1,\kappa,\kappa) \nn\\
&=
\label{squeezed_odd}
\frac{(-)^{\frac{n+3}{2}}\Ga\left(\tfrac{3}{2}\right)\kappa^n}{2^{\frac{3}{2}}(-v_1+v_2-v_3)\lm}
\sum_{\substack{j_2+j_3=\frac{n-3}{2} \\ j_2,j_3\geq 0}}
\frac{\Ga\left(-\tfrac{3}{2}-j_2\right)\Ga\left(n-\tfrac{3}{2}-j_3\right)}{j_2!j_3!}
+ O(\lm^0)\,.
\end{align}
Interestingly, by using the relation
\begin{equation}
\sum_{j=0}^m\frac{\Gamma(-\alpha-j)\Gamma(m+\alpha+j)}{\Gamma(j+1)\Gamma(m-j+1)}= -\frac{\pi}{\alpha\sin \pi\alpha} \delta_{m,0} \, ,
\end{equation}
it turns out that Eqs.~\eqref{squeezed_even} and~\eqref{squeezed_odd} are identically zero. Note that it is easy to see that the first nonzero contribution in the squeezed limit is $\mathcal{O}((\kappa_1/\kappa)^2)$. We therefore conclude that the triple-K integrals associated with CFT three-point functions of two nearly marginal scalars and one scalar with the conformal dimension close to a non-negative integer $n$ vanish in the squeezed limit, except for the case $n=0,3$, at the leading order in $\lambda$.

\end{document}